\documentclass[10pt, letter, onecolumn]{main}

\usepackage[pdftex]{graphicx}
\usepackage{pifont}
\usepackage{amssymb}
\usepackage{multirow}
\usepackage{array}
\usepackage{dblfloatfix}
\usepackage{titlesec}
\usepackage[numbers,sort&compress]{natbib}
\usepackage{nameref}
\usepackage{varioref}
\usepackage{etoc}

\newlength{\cboxlength}
\usepackage{amsmath,xparse,mleftright}
\NewDocumentCommand{\up}{som}{%
  \IfBooleanTF{#1}
    {\upext{#3}}
    {#3\IfNoValueTF{#2}{\mathord}{#2}\uparrow}%
}
\NewDocumentCommand{\upext}{m}{%
  \mleft.\kern-\nulldelimiterspace#1\mright\uparrow
}

\usepackage{xcolor} 
\usepackage{soul} 

\usepackage{adjustbox}
\usepackage[most]{tcolorbox}
\usepackage{float}
\usepackage{xspace}
\usepackage[symbol]{footmisc}
\usepackage{lineno}

\tcbset{
  aibox/.style={
    width=394.18663pt,
    top=10pt,
    colback=white,
    colframe=black,
    colbacktitle=black,
    enhanced,
    center,
    attach boxed title to top left={yshift=-0.1in,xshift=0.15in},
    boxed title style={boxrule=0pt,colframe=white,},
  }
}
\newtcolorbox{AIbox}[2][]{aibox,title=#2,#1}

\ifdefined\red
    \renewcommand{\red}[1]{\textcolor{red}{#1}}
\else
    \newcommand{\red}[1]{\textcolor{red}{#1}}
\fi
\ifdefined\blue
    \renewcommand{\blue}[1]{\textcolor{blue}{#1}}
\else
    \newcommand{\blue}[1]{\textcolor{blue}{#1}}
\fi

\newcommand{\name}[1][]{\textsc{CompRx}}

\makeatletter
\renewcommand\AB@authnote[1]{}
\renewcommand\AB@affilnote[1]{}
\makeatother

\usepackage{anyfontsize} 
\usepackage{titlesec}
\titleformat{\section}{\normalfont\Large\bfseries}{\thesection}{1em}{#1}

\title{{\fontsize{16.5pt}{15.5pt}\selectfont MedVAE: Efficient Automated Interpretation of Medical Images with Large-Scale Generalizable Autoencoders}}

\author[]{Maya Varma$^{1,*}$, Ashwin Kumar$^{1,*}$, Rogier van der Sluijs$^{1,*}$, Sophie Ostmeier$^{1}$, Louis Blankemeier$^{1}$, Pierre Chambon$^{1}$, Christian Bluethgen$^{1}$, Jip Prince$^{2}$, Curtis Langlotz$^{1}$, Akshay Chaudhari$^{1}$}

\affil{\footnotesize{$^1$Stanford Center for Artificial Intelligence in Medicine and Imaging, Stanford University, Palo Alto, CA, USA. $^2$UMC Utrecht, Utrecht, Netherlands}}

\renewcommand{\correspondingauthor}[1]{$\ast$~Equal contributions.}
\usepackage{xcolor}
\usepackage{hyperref}

\hypersetup{
    colorlinks=true,   
    urlcolor=blue      
}
\usepackage[noabbrev,capitalize]{cleveref}

\begin{document}
\begin{abstract}
Medical images are acquired at high resolutions with large fields of view in order to capture fine-grained features necessary for clinical decision-making. Consequently, training deep learning models on medical images can incur large computational costs. In this work, we address the challenge of downsizing medical images in order to improve downstream computational efficiency while preserving clinically-relevant features. We introduce \textit{MedVAE}, a family of six large-scale 2D and 3D autoencoders capable of encoding medical images as downsized latent representations and decoding latent representations back to high-resolution images. We train MedVAE autoencoders using a novel two-stage training approach with 1,052,730 medical images. Across diverse tasks obtained from 20 medical image datasets, we demonstrate that (1) utilizing MedVAE latent representations in place of high-resolution images when training downstream models can lead to efficiency benefits (up to 70x improvement in throughput) while simultaneously preserving clinically-relevant features and (2) MedVAE can decode latent representations back to high-resolution images with high fidelity. Our work demonstrates that large-scale, generalizable autoencoders can help address critical efficiency challenges in the medical domain. Our code is available at \href{https://github.com/StanfordMIMI/MedVAE}{https://github.com/StanfordMIMI/MedVAE}. 
\end{abstract}

\maketitle

\vspace{10mm}
\nolinenumbers
\clearpage
\section{Introduction}
Medical images (e.g. X-rays, computed tomography (CT) scans) are essential diagnostic tools in clinical practice. Since medical conditions are often characterized by the presence of subtle features, images are generally acquired with high spatial resolution and large fields of view in order to capture the required level of diagnostic detail for interpretation by radiologists~\cite{huda2015resolution}. However, high-resolution medical images, especially volumetric (3D) images, can result in large data storage costs and increased or even intractable computational complexity for downstream computer-aided diagnosis (CAD) models~\cite{freire2022computational, tan2019efficientnet}. This is likely to become a significant concern in the near future due to the rapid growth of medical imaging volumes stored by hospitals~\cite{mesterhazy2020high}, the expanding use of CAD tools in the clinic~\cite{engin2020,najjar2023}, and newer paradigm shifts towards large-scale foundation models~\cite{bommasani2022opportunities,chen2024chexagentfoundationmodelchest,merlin}. Many existing CAD models address this challenge by interpolating images to lower resolutions, despite the lower performance of models trained on interpolated data~\cite{sabottke2020effect, huang2023self}. 

A promising solution lies in powerful autoencoder methods, which are capable of encoding images as downsized latent representations and decoding latent representations back to images. Recent works, particularly in the context of latent diffusion models, have demonstrated that downsized latent representations can capture relevant spatial structure from high-resolution input images while simultaneously improving efficiency on tasks such as image generation~\cite{rombach2022high}. These findings suggest that autoencoders may hold potential for addressing the aforementioned storage and efficiency challenges in the medical domain by encoding high-resolution images as downsized latent representations, which can be used to develop downstream CAD models at a fraction of the computational cost.

Several large-scale autoencoders have been introduced in recent years~\cite{rombach2022high,lee2023llmcxr}; however, directly applying these models to the medical domain is challenging since medical images include a diverse range of clinically-relevant features (e.g. tumors, lesions, fractures), anatomical regions of focus (e.g. head, chest, knee), and modalities (e.g. 2D and 3D images). An effective generalizable autoencoding approach in the medical image domain must operate across a wide range of medical images and preserve clinically relevant features in both downsized latents as well as decoded reconstructions. However, existing autoencoder models are either (a) developed for natural images~\cite{rombach2022high}, which represent a significant domain shift from medical images, or (b) developed for a focused set of medical images (e.g. chest X-rays)~\cite{lee2023llmcxr} and are not explicitly trained to preserve clinically-relevant features across diverse medical images.

In this work, we address these limitations by introducing MedVAE, a family of 6 large-scale, generalizable 2D and 3D autoencoder models developed for the medical image domain. We first curate a large-scale training dataset with over one million 2D and 3D images, and we perform model training using a novel two-stage training scheme designed to optimize quality of latent representations and decoded reconstructions. 

We evaluate the quality of latent representations (using 8 CAD tasks) and reconstructed images (using both automated and manual perceptual quality evaluations) with respect to the preservation of clinically-relevant features. Evaluations are derived from 20 multi-institutional, open-source medical datasets with 4 imaging modalities  (X-ray, full-field digital mammograms, CT, and magnetic resonance imaging) and 8 anatomical regions. We measure the extent to which MedVAE latent representations and reconstructed images can contribute to downstream storage and efficiency benefits while simultaneously preserving clinically-relevant features. Ultimately, our results demonstrate that (1) downsized MedVAE latent representations can be used as drop-in replacements for high-resolution images in CAD pipelines while maintaining or exceeding performance; (2) downsized latent representations reduce storage requirements (up to 512x) and improve downstream efficiency of CAD model training (up to 70x in model throughput) when compared to high-resolution input images; and (3) decoded reconstructions effectively preserve clinically-relevant features as verified by an expert reader study. Our results also demonstrate that MedVAE models outperform existing natural image autoencoders. 

Ultimately, our work demonstrates the potential that large-scale, generalizable autoencoders hold in addressing the critical storage and efficiency challenges currently faced by the medical domain. Utilizing MedVAE latent representations instead of original, high-resolution images in training pipelines can improve model efficiency while preserving clinically-relevant features.

\clearpage
\section{Results}
\subsection{Training MedVAE autoencoders}

Autoencoding methods are capable of encoding high-resolution images as downsized latent representations. For a given 2D input image with dimensions $H \times W$ with $B$ channels, an autoencoding method will output a downsized latent representation of size $H/(\sqrt{f}) \times (W/\sqrt{f}) \times C$. Here, $f$ represents the downsizing factor applied to the 2D area of the image and $C$ represents a pre-specified number of latent channels. 3D autoencoding methods follow a similar formulation, where input images are 3D in nature with dimensions $H \times W \times S$ with $B$ channels. Here, the downsizing factor $f$ is applied to the 3D volume of the image; as a result, the latent representation will have dimensions $(H/(\sqrt[3]{f}) \times (W/\sqrt[3]{f}) \times (S/(\sqrt[3]{f}) \times C$. Autoencoding methods are also capable of decoding latent representations back to reconstructed high-resolution images. 

We aim to develop large-scale, generalizable medical image autoencoders capable of preserving diverse clinically-relevant features in both latent representations and reconstructions. To this end, we first collect a large-scale training dataset with 1,021,356 2D images and 31,374 3D images curated from 19 multi-institutional, open-source datasets~\cite{johnson2019mimic,feng2021candid,jeong2022emory,sorkhei2021csaw,rsnamammo,nguyen2022vindrmammo,moreira2012inbreast,cai2023online,jack2008alzheimer,dagley2017harvard,insel2020a4,lamontagne2019oasis,bien2018deep,hooper2021impact,chilamkurthy2018development,wasserthal2023totalsegmentator,ji2022amos,armato2011lung,stanfordaimi_coca_2024}. Images are obtained from two chest X-ray datasets, six full-field digital mammogram (FFDM) datasets, four T1- and T2-weighted head magnetic resonance imaging (MRI) datasets, one knee MRI dataset, two head/neck CT datasts, two whole-body CT datasets, and two chest CT datasets.

\begin{figure}[ht]
\centering
\includegraphics[width=\textwidth, trim=0 0 0 0]{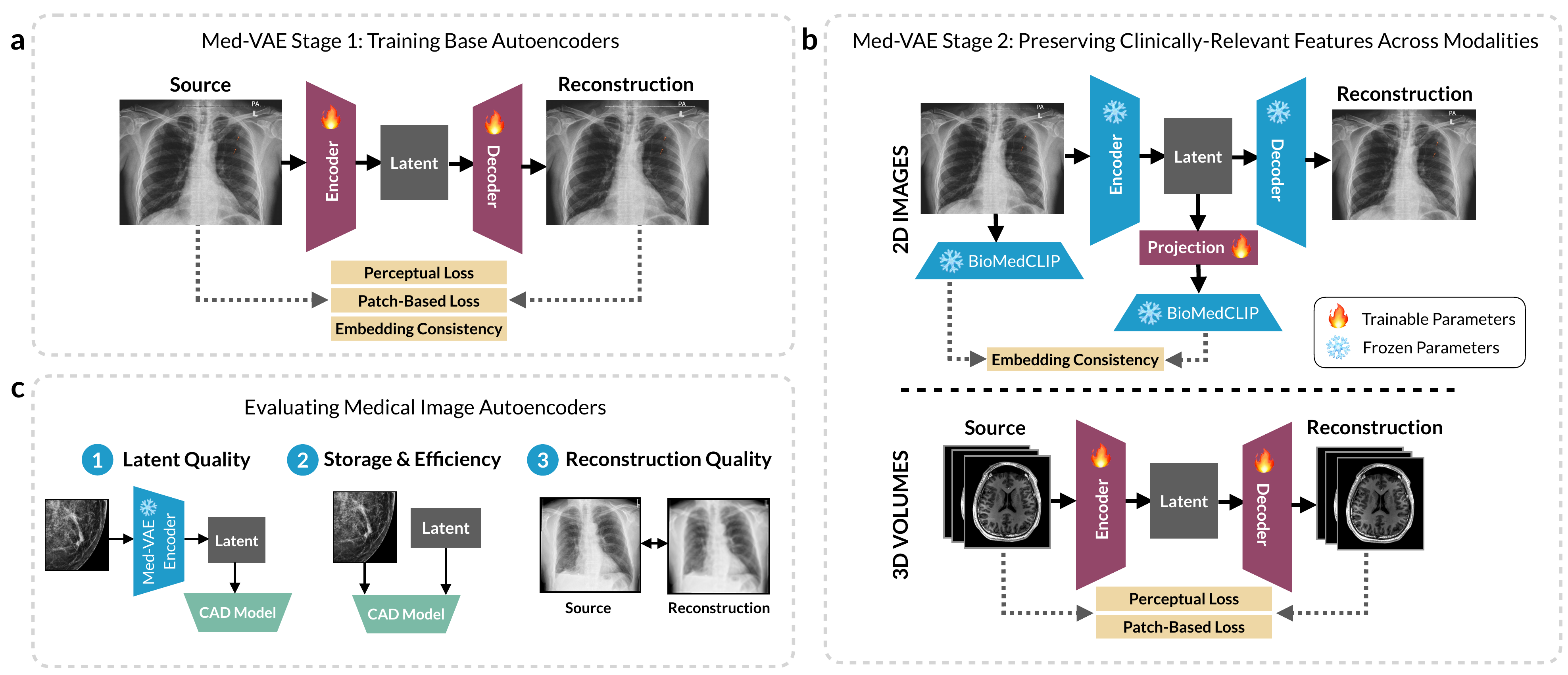}
\caption{\textbf{Overview of training pipeline and evaluation tasks for MedVAE, a suite of large-scale autoencoders for medical images.} \textbf{a,} We train MedVAE autoencoders using a two-stage process. The first stage involves training base autoencoders using 2D images. \textbf{b,} The second stage of training aims to further refine the latent space such that clinically-relevant features are preserved across modalities. We introduce separate training procedures for 2D images (e.g. X-rays, mammograms) and 3D volumes (e.g. CT scans, MRI). \textbf{c,} We evaluate medical image autoencoders with respect to latent representation quality, storage and efficiency benefits arising from using latent representations rather than high-resolution images in downstream CAD pipelines, and reconstructed image quality.}
\label{fig:method}
\end{figure}

We then utilize this dataset to train a family of generalizable autoencoders for medical images. Motivated by prior work on natural images~\cite{rombach2022high}, we utilize variational autoencoders (VAEs) as the model backbone. We perform model training using a novel two-stage training scheme designed to optimize quality of latent representations and decoded reconstructions. Specifically, the first stage involves training base autoencoders using 2D images (Fig.~\ref{fig:method}a); we maximize the perceptual similarity between input images and reconstructed images using a perceptual loss~\cite{lpips}, a patch-based adversarial objective~\cite{isola2018patchgan}, and a domain-specific embedding consistency loss. Whereas existing works on autoencoders train using only this stage, the medical image domain introduces the added complexity of subtle, fine-grained features required for clinical interpretation; thus, we introduce a second stage of training, which aims to further refine the latent space such that clinically-relevant features are preserved across various modalities (Fig.~\ref{fig:method}b). Specifically, in the context of 2D imaging modalities (e.g. X-rays, mammograms), the second training stage takes the form of a lightweight training approach that leverages the embedding space of BiomedCLIP, a recently-developed medical vision-language foundation model~\cite{zhang2023biomedclip}, to enforce feature consistency between input images and latent representations. In the context of 3D imaging modalities (e.g. CT scans, MRI), the second training stage involves lifting the 2D autoencoder architecture to 3D and performing continued fine-tuning with 3D images. In Appendix Table~\ref{table:ablations2d} and Appendix Table~\ref{table:ablations3d}, we analyze the effects of each stage of training on latent representation quality. In total, the MedVAE family includes 4 large-scale 2D autoencoders and 2 large-scale 3D autoencoders trained with various downsizing factors $f$ and latent channels $C$. 

In order to assess the capabilities of MedVAE, we evaluate the extent to which latent representations and reconstructed images generated by MedVAE autoencoders can contribute to downstream storage and efficiency benefits while simultaneously preserving clinically-relevant features (Fig.~\ref{fig:method}c). Specifically, we assess (1) whether downsized latent representations can effectively replace high-resolution images in CAD pipelines while maintaining performance; (2) whether latent representations can reduce storage requirements and improve downstream efficiency; and (3) whether decoded reconstructions effectively preserve clinically-relevant features necessary for radiologist interpretation. 

\subsection{Latent representation quality}

\begin{table*}[t]

\centering
\resizebox{\linewidth}{!}{%
{
\begin{tabular}{ lcccccccc }
\toprule
\textbf{}
& \multicolumn{2}{c}{\textbf{}}
& \multicolumn{5}{c}{\textbf{AUROC} $\uparrow$}
& \textbf{}
\\
\cmidrule(l{3pt}r{0pt}){4-8}
\cmidrule(l{3pt}r{0pt}){9-9}
\textbf{Method}
& \textbf{$f$}
& \textbf{$C$}
& \small Malignancy
& \small Calcification
& \small BI-RADS
& \small Bone Age
& \small Wrist Fracture
& Average
\\ 
 & & &  \small (FFDM) & \small  (FFDM) & \small  (FFDM) &  (X-ray) &  (X-ray)
\\
\midrule
\small High-Resolution & 1 & 1 & \textbf{66.1$_{\pm0.5}$} & \textbf{62.4$_{\pm0.6}$} & \textbf{63.4$_{\pm0.1}$} & \textbf{80.2$_{\pm0.1}$}  & \textbf{73.7$_{\pm0.0}$} & \textbf{69.2}\\

\midrule
\small Nearest & 16 & 1   &  65.5$_{\pm0.1}$ & 59.7$_{\pm0.3}$ & 62.4$_{\pm0.1}$ & \textcolor{blue}{81.6$_{\pm0.1}$}  & 70.5$_{\pm0.0}$ & 67.9\\
\small Bilinear & 16 & 1    &65.5$_{\pm0.1}$ & 58.1$_{\pm0.3}$ & 61.1$_{\pm0.2}$ & \textcolor{blue}{81.6$_{\pm0.0}$}  & \textbf{71.2$_{\pm0.1}$} & 67.5 \\
\small Bicubic & 16 & 1   &  65.5$_{\pm0.4}$ & 58.5$_{\pm0.5}$ & 61.1$_{\pm0.0}$ & \textcolor{blue}{81.8$_{\pm0.2}$}  & 71.1$_{\pm0.1}$ & 67.6\\
\small KL-VAE & 16 & 3   & 59.7$_{\pm0.2}$ & 59.1$_{\pm0.3}$ & 58.5$_{\pm0.1}$ & 74.3$_{\pm0.1}$ & 64.5$_{\pm0.1}$ & 63.2\\
\small VQ-GAN & 16 & 3   & 57.4$_{\pm0.3}$  & 58.2$_{\pm0.4}$ & 62.3$_{\pm0.1}$ & 79.1$_{\pm0.2}$ & 65.8$_{\pm0.1}$ & 64.6\\
\small 2D MedVAE & 16 & 1   &  63.6$_{\pm0.6}$ & \textcolor{blue}{\textbf{63.9$_{\pm0.4}$}} & \textcolor{blue}{\textbf{65.3$_{\pm0.2}$}} & \textcolor{blue}{\textbf{84.6$_{\pm0.1}$}} & 70.3$_{\pm0.1}$ & \textcolor{blue}{\textbf{69.5}}\\
\small 2D MedVAE & 16  & 3 &  \textcolor{blue}{\textbf{66.1$_{\pm0.2}$}} &   61.7$_{\pm0.2}$ & 62.3$_{\pm0.1}$ & \textcolor{blue}{82.1$_{\pm0.1}$}  &   70.6$_{\pm0.1}$ & 68.6\\
\midrule

\small Nearest & 64 & 1    & 63.0$_{\pm0.1}$ & 58.8$_{\pm0.2}$ & 60.0$_{\pm0.2}$ & 72.1$_{\pm0.0}$  & 65.1$_{\pm0.1}$ & 63.8	\\
\small Bilinear & 64 & 1   & 61.5$_{\pm0.3}$ & 56.9$_{\pm0.4}$ & \textbf{61.3$_{\pm0.1}$} & 72.8$_{\pm0.5}$  & \textbf{67.9$_{\pm0.1}$} & 64.1\\
\small Bicubic & 64 & 1   & 61.2$_{\pm0.5}$ & 57.6$_{\pm0.4}$ & 61.1$_{\pm0.1}$ & 72.8$_{\pm0.2}$  & 67.9$_{\pm0.2}$ & 64.1\\
\small KL-VAE & 64 & 4   & 62.2$_{\pm0.7}$ &  55.8$_{\pm0.4}$ & 56.8$_{\pm0.1}$ & 65.7$_{\pm0.0}$ & 58.8$_{\pm0.0}$ & 59.9\\
\small VQ-GAN & 64 & 4    & 64.5$_{\pm0.5}$ & 57.3$_{\pm0.3}$ &  56.6$_{\pm0.1}$ & 67.6$_{\pm0.1}$  & 61.6$_{\pm0.2}$ & 61.5 \\
\small 2D MedVAE & 64 & 1  & 59.0$_{\pm0.3}$ & \textbf{59.4$_{\pm0.7}$} & 60.7$_{\pm0.1}$ & \textbf{73.5$_{\pm0.2}$} & 64.3$_{\pm0.1}$ & 63.4\\
\small 2D MedVAE & 64 & 4   & \textbf{64.9$_{\pm0.2}$} &  58.5$_{\pm0.3}$ & 60.6$_{\pm0.0}$ & 73.0$_{\pm0.2}$ & 66.7$_{\pm0.1}$ & \textbf{64.7}\\

\bottomrule
\end{tabular}
}
}
\caption{\textbf{Evaluating latent representation quality with 2D CAD tasks.} We evaluate the 2D MedVAE autoencoders on five 2D CAD tasks, and we report the mean AUROC and standard deviation across three random seeds. We compare MedVAE with three interpolation methods (nearest, bilinear, bicubic) and two natural image autoencoders (KL-VAE and VQ-GAN). Here, $f$ represents the downsizing factor applied to the 2D area of the input image and $C$ represents the number of latent channels. The best performing models on each task are bolded. We highlight methods that perfectly preserve clinically-relevant features (i.e. performance equals or exceeds performance when training with high-resolution images) in \textcolor{blue}{\textbf{blue}}.}
\label{table:image_classification}
\vspace{-1mm}
\end{table*}

\begin{table*}[ht]
\centering
{%
\begin{tabular}{ lcccccc }
\toprule
\textbf{}
& \multicolumn{2}{c}{\textbf{}}
& \multicolumn{3}{c}{\textbf{AUROC} $\uparrow$}
& \textbf{}
\\
\cmidrule(l{3pt}r{0pt}){4-6}
\cmidrule(l{3pt}r{0pt}){7-7}
\textbf{Method}
& \textbf{$f$}
& \textbf{$C$}
& \small Spine Fractures
& \small Skull Fractures
& \small Knee Injury
& Average
\\ 
 & & &  \small (CT) & \small (CT) & \small (MRI) &
\\
\midrule
\small High-Resolution & 1 & 1  & \textbf{82.9$_{\pm2.2}$} & \textbf{63.9$_{\pm6.3}$} & \textbf{69.9$_{\pm0.6}$} & \textbf{72.2}\\

\midrule
\small Bicubic & 64 & 1  &  77.3$_{\pm4.1}$ & \textcolor{blue}{64.8$_{\pm4.0}$} & 66.4$_{\pm2.3}$ & 69.5\\
\small KL-VAE & 64 & 3   & 68.8$_{\pm2.1}$ & 40.7$_{\pm9.1}$ & 63.9$_{\pm8.2}$  &  57.8\\
\small VQ-GAN & 64 & 3 & 73.2$_{\pm2.0}$  & 75.5$_{\pm14.8}$ & 63.6$_{\pm10.5}$ &   70.8 \\
\small 3D MedVAE & 64 & 1  & \textcolor{blue}{\textbf{83.7$_{\pm2.8}$}} & \textcolor{blue}{\textbf{87.0$_{\pm7.3}$}} & \textbf{68.4$_{\pm2.4}$} & \textcolor{blue}{\textbf{79.7}}\\
\midrule
\small Bicubic & 512 & 1 & \textbf{72.3$_{\pm2.2}$} & 38.4$_{\pm24.5}$ & \textbf{59.4$_{\pm2.5}$} & 56.7\\
\small KL-VAE & 512 & 4  & 67.7$_{\pm3.9}$ & 42.6$_{\pm4.0}$ & 50.9$_{\pm5.1}$ & 53.7\\
\small VQ-GAN & 512 & 4  & 68.9$_{\pm7.0}$ & 30.6$_{\pm12.5}$ & 57.4$_{\pm5.0}$ & 52.3 \\
\small 3D MedVAE & 512 & 1  & 72.0$_{\pm3.8}$ & \textbf{49.1$_{\pm19.8}$} & 58.2$_{\pm1.7}$ & \textbf{59.8} \\
\bottomrule
\end{tabular}
}
\caption{\textbf{Evaluating latent representation quality with 3D CAD tasks.} We evaluate the 3D MedVAE autoencoders on three 3D CAD tasks, and we report the mean AUROC and standard deviation across three random seeds. We compare MedVAE with one interpolation method (bicubic) and two natural image 2D autoencoders (KL-VAE and VQ-GAN). For 2D baselines, we stitch 2D latent representations together across slices such that the size of the 2D latent representation matches those generated by 3D models. Here, $f$ represents the downsizing factor applied to the 3D volume of the input image and $C$ represents the number of latent channels. The best performing models on each task are bolded. We highlight methods that perfectly preserve clinically-relevant features (i.e. performance equals or exceeds performance when training with high-resolution volumes) in \textcolor{blue}{\textbf{blue}}.}
\label{table:3D_latent_cls}
\vspace{-1mm}
\end{table*}

We first evaluate whether clinically-relevant features are preserved in MedVAE latent representations. To this end, we measure the extent to which latent representations can serve as drop-in replacements for high-resolution input images in CAD pipelines \textit{without} any customization or modifications to CAD model architectures. 

We evaluate latent representation quality using the following 8 CAD tasks: malignancy detection on 2D FFDMs~\cite{cai2023online}, calcification detection on 2D FFDMs~\cite{cai2023online}, BI-RADS prediction on 2D FFDMs~\cite{nguyen2022vindrmammo}, bone age prediction on 2D X-rays~\cite{rsnaboneage}, fracture detection on 2D wrist X-rays~\cite{Nagy2022wristfrac}, fracture detection on 3D spine CTs~\cite{loffler2020vertebral}, fracture classification on 3D head CTs~\cite{chilamkurthy2018development}, and anterior cruciate liagment (ACL) and meniscal tear detection on 3D sagittal knee MRIs~\cite{bien2018deep}. In order to perform each of these CAD tasks, a model must rely on fine-grained, clinically-relevant features.

For each CAD task, we train a classifier (HRNet~\cite{wang2020hrnet} in 2D settings and SEResNet~\cite{hu2018squeeze} in 3D settings) on a training set consisting of latent representations. We then measure the difference in classification performance between models trained directly on latent representations and models trained using original, high-resolution images; this serves as an indicator of latent representation quality (e.g. a small performance difference indicates that the downsizing approach preserves diagnostic features). We compute AUROC for all binary tasks and macro AUROC for all multi-class tasks. We train each classifier with three random seeds, and we report results as mean AUROC $\pm$ standard deviation.

We compare MedVAE with two categories of image downsizing methods: (1) interpolation methods (nearest, bilinear, and bicubic), which are the de-facto gold standard for medical image downsizing as demonstrated by the quantity of prior work leveraging this approach~\cite{wantlin2023benchmd, Varma2019, convirt, Huang_2021_ICCV, miura2021improving, Tiu2022}, and (2) recently-introduced large-scale natural image autoencoders (KL-VAE and VQ-GAN)~\cite{rombach2022high}. Due to the fact that prior work on developing large-scale 3D autoencoders has been limited, we compare our 3D MedVAE models with 2D methods by stitching 2D latent representations together across slices such that the size of the 2D latent representation matches those generated by 3D models. 

We provide results for 2D CAD tasks in Table \ref{table:image_classification} and 3D CAD tasks in Table \ref{table:3D_latent_cls}. Our results demonstrate that the MedVAE training approach yields high-quality latent representations for both 2D and 3D images. At a downsizing factor of $f=16$, 2D MedVAE perfectly preserves clinically-relevant features on four out of five 2D classification tasks. Similarly, at a downsizing factor of $f=64$, 3D MedVAE perfectly preserves relevant clinical information on two out of three 3D classification tasks (spine and skull CT fracture detection). In these cases, performance equals or exceeds performance when training with original, high-resolution images. The average performance of 2D MedVAE with $f=16$ and 3D MedVAE with $f=64$ across all tasks also exceeds the average performance when training with high-resolution images.  

We observe that MedVAE consistently outperforms the natural image autoencoders KL-VAE and VQ-GAN on all classification tasks, demonstrating the utility of the MedVAE training procedure. On average across the five 2D classification tasks, 2D MedVAE demonstrates a 10.0\% improvement over KL-VAE at a downsizing factor of $f=16$ and a 8.0\% improvement at a downsizing factor of $f=64$. Similar trends are noted for VQ-GAN. 
In particular, 2D MedVAE outperforms KL-VAE and VQ-GAN on the two musculoskeletal tasks (bone age prediction and wrist fracture detection) despite the fact that no musculoskeletal radiographs are used during MedVAE training; this suggests effective generalization to other types of medical images. On average across the three 3D tasks, 3D MedVAE demonstrates a 37.9\% improvement over KL-VAE at a downsizing factor of $f=64$ and a 11.4\% improvement over KL-VAE at a downsizing factor of $f=512$. Our findings suggest that 3D training of autoencoders leads to high-quality latent representations due to preservation of volumetric information (e.g. fractures spanning multiple slices), particularly at $f=64$. These findings are corroborated by results in Appendix Table~\ref{table:2d3dcad}, which compares performance of 2D MedVAE and 3D MedVAE on 3D CAD tasks.

Additionally, we note that MedVAE outperforms the interpolation methods across most tasks, but interpolation methods are a competitive baseline. Overall, our findings suggest that our MedVAE training procedure yields downsized latent representations that can be used as drop-in replacements for high-resolution input images in CAD pipelines. 

\subsection{Storage and efficiency benefits of latent representations}

Next, we evaluate the extent to which downsized MedVAE latent representations can reduce storage requirements and improve downstream efficiency of CAD pipelines when compared to high-resolution input images. Using a 2D high-resolution network (HRNet\_w64) and 3D squeeze-excitation network (SEResNet-152) as our base CAD architectures, we report latency, throughput, and maximum batch size. Latency is the time (in milliseconds) to perform a forward pass of the network on one batch. Throughput is the number of samples that can be evaluated by the network in one second. Finally, we report the maximum batch size (in powers of 2) for a forward pass that will fit on a single A100 GPU (2D) and an A6000 GPU (3D). We assume a high-resolution input image size of $1024 \times 1024$ with 1 channel for 2D settings and an input volume size of $256 \times 256 \times 256$ with 1 channel for 3D settings.

\begin{figure}[ht]
\centering
\includegraphics[width=\textwidth, trim=0 0 0 0]{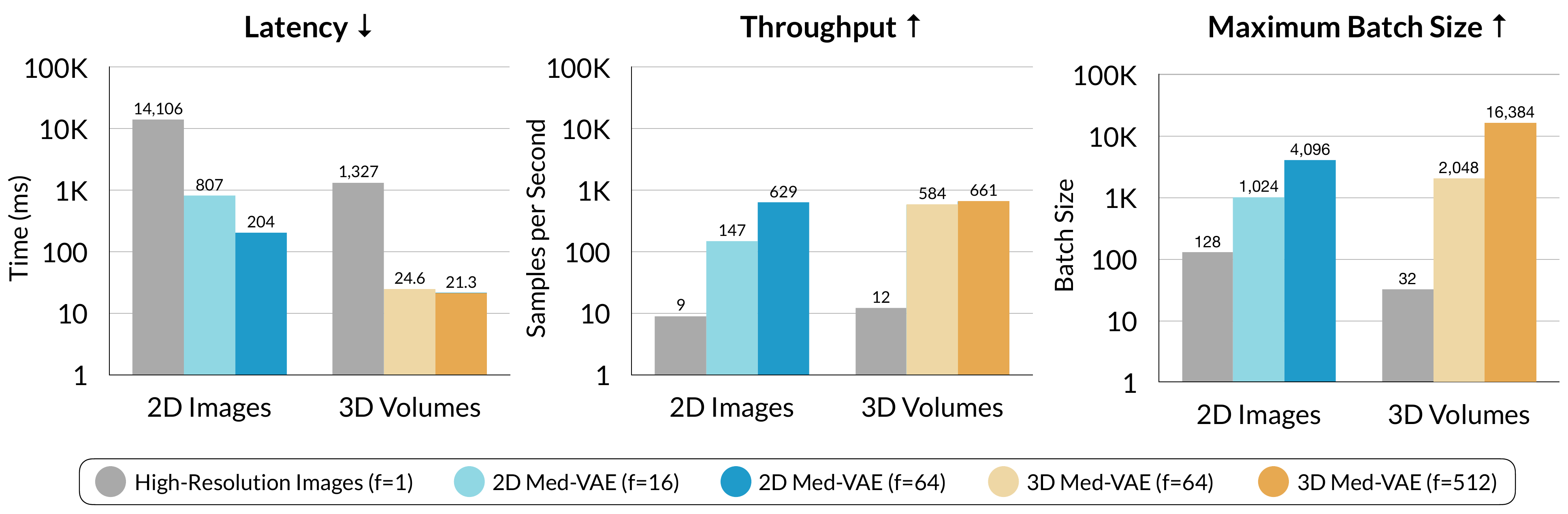}
\caption{\textbf{CAD model efficiency.} Here, we compare the efficiency of CAD models trained with downsized latent representations to CAD models trained with high-resolution images. $f$ represents the downsizing factor applied to the 2D area or 3D volume of the input image. We report latency (in milliseconds), throughput (in samples per second), and the maximum batch size (in powers of 2) that will fit on one GPU.}
\label{fig:efficiency}
\end{figure}

Results are provided in Figure \ref{fig:efficiency}. We demonstrate that training CAD models directly on downsized latent representations can lead to large improvements in model efficiency. In the 2D setting, we observe that as the downsizing factor increases to $f=64$, the latency decreases by 69x, the throughput increases by 70x, and the maximum batch size increases by 32x. In the 3D setting, as the downsizing factor increases to $f=512$, the latency decreases by 62x, the throughput increases by 55x, and the maximum batch size increases by 512x. Storage costs decrease proportionally with the downsizing factor (i.e. 64x for 2D and 512x for 3D).

\subsection{Reconstructed image quality}

We evaluate whether clinically-relevant features are preserved in reconstructed images using both automated and manual perceptual quality evaluations. These evaluations quantify the extent to which the encoding and subsequent decoding processes retain relevant features.

For automated evaluations, we use perceptual metrics to compare reconstructed images with the original inputs. We report peak signal-to-noise ratio (PSNR) and the multi-scale structural similarity index measure (MS-SSIM). For 2D evaluations, we measure perceptual quality on X-rays~\cite{feng2021candid,johnson2019mimic}; FFDMs~\cite{jeong2022emory,sorkhei2021csaw,rsnamammo,nguyen2022vindrmammo,moreira2012inbreast,cai2023online}; and musculoskeletal X-rays~\cite{Nagy2022wristfrac}. In addition to full-image evaluations, we additionally include a fine-grained perceptual quality assessment, where we extract regions containing wrist fractures by using bounding boxes~\cite{Nagy2022wristfrac}; then, the original region and reconstructed region are compared using perceptual metrics. For 3D evaluations, we compute metrics on brain MRIs~\cite{jack2008alzheimer,dagley2017harvard,insel2020a4,lamontagne2019oasis}; head CTs~\cite{chilamkurthy2018development}; abdomen CTs~\cite{ji2022amos}; CTs from a wide range of anatomies~\cite{wasserthal2023totalsegmentator}; lung CTs~\cite{armato2011lung}; and knee MRIs~\cite{bien2018deep}.

\begin{table*}[t]
\centering
\resizebox{\linewidth}{!}{
\begin{tabular}{lccccccccc}
\toprule
\textbf{Method}
& \textbf{$f$}
& \textbf{$C$}
& \multicolumn{2}{c}{\textbf{Mammograms}}
& \multicolumn{2}{c}{\textbf{Chest X-rays}}
& \multicolumn{2}{c}{\textbf{Musculoskeletal X-rays}}
& \multicolumn{1}{c}{\textbf{Wrist X-rays (FG)}}
\\
\cmidrule(l{3pt}r{0pt}){4-5}
\cmidrule(l{3pt}r{0pt}){6-7}
\cmidrule(l{3pt}r{0pt}){8-9}
\cmidrule(l{3pt}r{0pt}){10-10}
& 
&
& \small PSNR $\uparrow$
& \small MS-SSIM $\uparrow$
& \small PSNR $\uparrow$
& \small MS-SSIM $\uparrow$
& \small PSNR $\uparrow$
& \small MS-SSIM $\uparrow$
& \small PSNR $\uparrow$
\\ 
\midrule
\small Nearest & 16 & 1    &  25.95$_{\pm0.06}$  & 0.846$_{\pm0.00}$  & 29.87$_{\pm0.04}$ & 0.942$_{\pm0.00}$   &  24.06$_{\pm0.02}$  & 0.890$_{\pm0.00}$  & 26.11$_{\pm0.02}$\\
\small Bilinear & 16 & 1    & 30.18$_{\pm0.07}$  & 0.936$_{\pm0.00}$ & 34.23$_{\pm0.03}$ & 0.981$_{\pm0.00}$ & 28.75$_{\pm0.02}$  & 0.959$_{\pm0.00}$ & 30.92$_{\pm0.03}$ \\
\small Bicubic & 16 & 1    & 31.69$_{\pm0.07}$  & 0.961$_{\pm0.00}$ & 35.48$_{\pm0.03}$ & 0.989$_{\pm0.00}$ & 30.18$_{\pm0.02}$  & 0.974$_{\pm0.00}$ & 32.65$_{\pm0.04}$  \\
\small KL-VAE & 16 & 3  & 36.11$_{\pm0.07}$  & 0.989$_{\pm0.00}$ & 41.45$_{\pm0.04}$ & 0.996$_{\pm0.00}$ & 38.29$_{\pm0.03}$  &0.992$_{\pm0.00}$  & 36.55$_{\pm0.03}$\\
\small VQ-GAN & 16 & 3  &   35.55$_{\pm0.07}$  & 0.986$_{\pm0.00}$ & 37.80$_{\pm0.03}$ & 0.995$_{\pm0.00}$ & 36.41$_{\pm0.02}$  & 0.990$_{\pm0.00}$  & 34.19$_{\pm0.04}$\\
\small 2D MedVAE  & 16 & 1  & 32.34$_{\pm0.07}$ &  0.969$_{\pm0.00}$ & 38.44$_{\pm0.02}$ & 0.990$_{\pm0.00}$ & 33.97$_{\pm0.03}$ & 0.973$_{\pm0.00}$  &  31.97$_{\pm0.03}$ \\
\small 2D MedVAE  & 16 & 3  & \textbf{37.57}$_{\pm0.08}$ &  \textbf{0.993}$_{\pm0.00}$ & \textbf{43.55 }$_{\pm0.02}$& \textbf{0.997}$_{\pm0.00}$ & \textbf{39.41}$_{\pm0.04}$ & \textbf{0.994}$_{\pm0.00}$  &  \textbf{37.61}$_{\pm0.02}$  \\
\midrule

\small Nearest & 64 & 1   & 22.46$_{\pm0.05}$  & 0.669$_{\pm0.00}$ & 26.22$_{\pm0.03}$ & 0.858$_{\pm0.00}$ & 19.93$_{\pm0.02}$  & 0.756$_{\pm0.00}$  & 22.14$_{\pm0.04}$  \\
\small Bilinear & 64 & 1    & 26.81$_{\pm0.06}$  & 0.837$_{\pm0.00}$ & 31.18$_{\pm0.03}$ & 0.949$_{\pm0.00}$ & 24.89$_{\pm0.01}$  & 0.898$_{\pm0.00}$ & 27.12$_{\pm0.03}$  \\
\small Bicubic & 64 & 1    & 27.84$_{\pm0.06}$  & 0.874$_{\pm0.00}$ & 32.09$_{\pm0.03}$ & 0.962$_{\pm0.00}$  & 25.92$_{\pm0.01}$  & 0.922$_{\pm0.00}$  & 28.54$_{\pm0.03}$ \\
\small KL-VAE & 64 & 4    & 31.88$_{\pm0.07}$  & 0.959$_{\pm0.00}$ &36.37$_{\pm0.01}$ &0.987$_{\pm0.00}$ &33.49$_{\pm0.02}$ &0.966$_{\pm0.00}$ & 31.04$_{\pm0.03}$ \\
\small VQ-GAN & 64 & 4   & 30.13$_{\pm0.06}$  & 0.938$_{\pm0.00}$ & 34.87$_{\pm0.02}$ & 0.980$_{\pm0.00}$ & 32.00$_{\pm0.02}$  & 0.953$_{\pm0.0}$  & 29.92$_{\pm0.02}$ \\
\small 2D MedVAE  & 64 & 1   & 28.00$_{\pm0.07}$ & 0.872$_{\pm0.00}$ & 31.92$_{\pm0.04}$ &  0.962$_{\pm0.00}$ & 28.27$_{\pm0.02}$&  0.917$_{\pm0.00}$  & 28.03$_{\pm0.01}$\\
\small 2D MedVAE  & 64 & 4  & \textbf{33.13}$_{\pm0.07}$  & \textbf{0.969}$_{\pm0.00}$ & \textbf{38.88}$_{\pm0.03}$ &  \textbf{0.990}$_{\pm0.00}$ & \textbf{34.73}$_{\pm0.02}$&  \textbf{0.972}$_{\pm0.00}$& \textbf{32.30}$_{\pm0.02}$\\
\bottomrule
\end{tabular}
}
\caption{\textit{Evaluating reconstruction quality on 2D datasets.} We evaluate 2D MedVAE with perceptual quality metrics on mammograms and chest X-rays, which we classify as \textit{in-distribution}, since the MedVAE training set includes mammograms and chest X-rays. We also evaluate MedVAE on musculoskeletal X-rays and wrist X-rays (fine-grained), which we classify as \textit{out-of-distribution}. Here, $f$ represents the downsizing factor applied to the 2D area of the input image and $C$ represents the number of latent channels. The best performing models are bolded. We calculate PSNR and MS-SSIM using a random sample of 1000 images for each image type; we report mean and standard deviations across four runs with different random seeds.}
\label{table:perceptualid}
\end{table*}

\begin{table*}[t]
\centering
\resizebox{\linewidth}{!}{
\begin{tabular}{lcccccccccccccc}
\toprule
\textbf{Method}
& \textbf{$f$}
& \textbf{$C$}
& \multicolumn{2}{c}{\textbf{Brain MRIs}}
& \multicolumn{2}{c}{\textbf{Head CTs}}
& \multicolumn{2}{c}{\textbf{Abdomen CTs}}
& \multicolumn{2}{c}{\textbf{TS CTs}}
& \multicolumn{2}{c}{\textbf{Lung CTs}}
& \multicolumn{2}{c}{\textbf{Knee MRIs}}
\\
\cmidrule(l{3pt}r{0pt}){4-5}
\cmidrule(l{3pt}r{0pt}){6-7}
\cmidrule(l{3pt}r{0pt}){8-9}
\cmidrule(l{3pt}r{0pt}){10-11}
\cmidrule(l{3pt}r{0pt}){12-13}
\cmidrule(l{3pt}r{0pt}){14-15}
& 
&
& \small PSNR $\uparrow$
& \small MS-SSIM $\uparrow$
& \small PSNR $\uparrow$
& \small MS-SSIM $\uparrow$
& \small PSNR $\uparrow$
& \small MS-SSIM $\uparrow$
& \small PSNR $\uparrow$
& \small MS-SSIM $\uparrow$
& \small PSNR $\uparrow$
& \small MS-SSIM $\uparrow$
& \small PSNR $\uparrow$
& \small MS-SSIM $\uparrow$
\\ 
\midrule
\small Bicubic & 16 & 1 & 29.27 & 0.975 & 36.21 & 0.996 & 33.81 & 0.989 & 27.33 & 0.972 & 28.00 & 0.973 & 26.37 & 0.986 \\
\small KL-VAE & 16 & 3 & 33.23 & {\textbf{0.994}} & 47.65 & {\textbf{1.000}} & 43.51 & 0.998 & 34.14 & 0.994 & 32.62 & {\textbf{0.989}} & 31.31 & {\textbf{0.998}} \\
\small VQ-GAN & 16 & 3 & 32.72 & 0.992 & 42.87 & 0.999 & 40.85 & 0.997 & 33.55 & 0.993 & 32.20 & {\textbf{0.989}} & 30.75 & 0.997 \\
\small 2D MedVAE  & 16 & 1 & 29.48 & 0.980 & 39.71 & 0.997 & 33.45 & 0.983 & 29.70 & 0.983 & 28.40 & 0.973 & 27.38 & 0.990 \\
\small 2D MedVAE  & 16 & 3 & {\textbf{33.99}} & {\textbf{0.994}} & {\textbf{48.56}} & {\textbf{1.000}} & {\textbf{44.95}} & {\textbf{0.999}} & {\textbf{34.83}} & {\textbf{0.995}} & {\textbf{33.34}} & {\textbf{0.989}} & {\textbf{31.52}} & 0.997 \\
\small 3D MedVAE  & 64 & 1 & 29.52 & 0.983 & 39.03 & 0.999 & 36.61 & 0.993 & 31.35 & 0.987 & 28.79 & 0.975 & 28.25 & 0.994 \\
\midrule
\small Bicubic & 64 & 1 & 26.25 & 0.911 & 30.11 & 0.980 & 28.84 & 0.955  & 24.24 & 0.914 & 24.40 & 0.928 & 24.11 & 0.956  \\
\small KL-VAE & 64 & 3 & 29.32 & {\textbf{0.977}} & 40.95 & 0.997 & 38.07 & {\textbf{0.995}} & 29.85 & 0.982 & 28.83 & 0.974 & 27.68 & {\textbf{0.993}} \\
\small VQ-GAN & 64 & 3 & 27.43 & 0.967 & 39.02 & 0.997 & 36.25 & 0.991 & 27.47 & 0.972 & 26.66 & 0.964 & 25.95 & 0.990 \\
\small 2D MedVAE  & 64 & 1 & 25.66 & 0.920 & 33.10 & 0.988 & 29.51 & 0.967  & 24.50 & 0.922 & 24.39 & 0.933 & 24.48 & 0.973 \\
\small 2D MedVAE  & 64 & 3 & {\textbf{29.34}} & 0.976 & {\textbf{41.98}} & {\textbf{0.999}} & {\textbf{39.49}} & {\textbf{0.995}} & {\textbf{30.35}} & {\textbf{0.984}} & {\textbf{29.59}} & {\textbf{0.977}} & {\textbf{28.05}} & {\textbf{0.993}} \\
\small 3D MedVAE  & 512 & 1 & 26.23 & 0.937 & 30.85 & 0.991 & 29.47 & 0.960 & 26.34 & 0.949 & 24.76 & 0.934 & 24.36 & 0.977 \\

\bottomrule
\end{tabular}
}
\caption{\textit{Evaluating reconstruction quality on 3D datasets.} We evaluate 3D MedVAE with perceptual quality metrics on head MRIs, head CTs, abdomen CTs, various high-resolution CTs (TS), lung CTs, and knee MRIs. $f$ represents the downsizing factor applied to the input volume and $C$ represents the number of latent channels. The best performing models are bolded. We compare 3D MedVAE with several 2D methods, including 2D MedVAE, KL-VAE, and VQ-GAN.}
\label{table:3dperceptual}
\end{table*}

In Table \ref{table:perceptualid}, we compare 2D MedVAE with interpolation methods and large-scale natural image autoencoders across four types of 2D images. We find that 2D MedVAE achieves the highest perceptual quality across all evaluated image types. In particular, our evaluations with wrist X-rays explore generalization of MedVAE to unseen anatomical features; notably, MedVAE achieves the highest PSNR scores on this task, despite the fact that MedVAE was not trained on musculoskeletal X-rays. We also note a general trend that increasing the number of latent channels $C$ improves perceptual quality of the reconstructed image. 

In Table \ref{table:3dperceptual}, we compare MedVAE with interpolation methods and large-scale natural image autoencoders across six types of 3D volumes. Due to the absence of existing large-scale 3D autoencoder baselines, we compare our 3D MedVAE models with 2D methods by performing downsizing on individual 2D slices and then stitching slices together to form the reconstructed 3D volume. We again find that MedVAE reconstructions demonstrate superior perceptual quality when compared to baselines. In particular, 2D MedVAE achieves the highest perceptual quality across almost all evaluated image types, despite the fact that no MRI or CT slices were included in the 2D MedVAE training set. We also observe that 3D MedVAE achieves competitive performance, despite utilizing a significantly higher downsizing factor than comparable 2D methods (i.e. downsizing across all three dimensions rather than just two). In Appendix Table~\ref{table:decoder}, we compare 3D MedVAE with a model referred to as 2D MedVAE-Decoder, which has a comparable downsizing factor $f$. The 2D MedVAE-Decoder model performs downsizing on individual 2D slices, which are then stitched and interpolated together to form a latent representation of equivalent size to the 3D MedVAE model; we then perform fine-tuning of the decoder using our curated dataset of 3D volumes. The superiority of 3D MedVAE to the 2D MedVAE-Decoder approach demonstrates the utility of 3D training of autoencoders, which enables the model to capture important volumetric patterns. 

\begin{figure}[h]
\centering
\includegraphics[width=0.9\textwidth, trim=0 0 0 0]{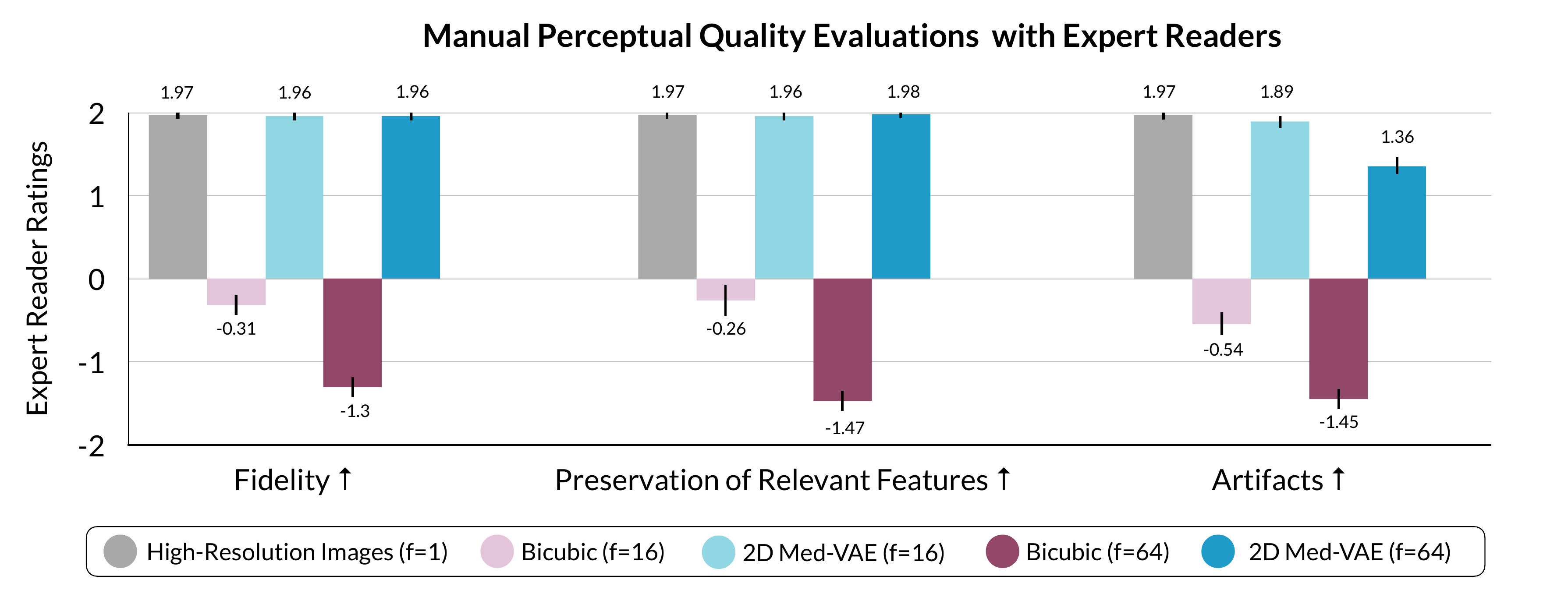}
\caption{\textbf{Manual perceptual quality evaluations with expert readers.} We report the mean scores from three expert readers on three criteria: fidelity, preservation of relevant features, and artifacts. We compare 2D MedVAE with ($f=16,C=3$) and ($f=64,C=4$) with bicubic interpolation, a standard and widely-used approach for downsizing medical images. Error bars represent 95\% confidence intervals.}
\label{fig:readerstudy}
\end{figure}

Qualitative reader studies by domain experts are critical for ensuring clinical usability of developed methods. We supplement our automated evaluations of reconstructed image quality with a manual reader study. Each reader is presented with a pair of chest X-rays, consisting of an original high-resolution image on the left and a reconstructed image on the right. A total of 50 unique chest X-rays with fractures, randomly sampled from CANDID-PTX, are selected and presented in a randomized order~\cite{feng2021candid}. The reconstructed images are scored on a 5-point Likert scale ranging from -2 to 2 based on three main criteria: image fidelity, preservation of diagnostic features, and the presence of artifacts. Our study involved three radiologists as expert readers. We compared 2D MedVAE with bicubic interpolation, a standard and widely-used approach for downsizing medical images (Figure~\ref{fig:readerstudy}).

\begin{figure}[h]
\centering
\includegraphics[width=0.9\textwidth, trim=0 0 0 0]{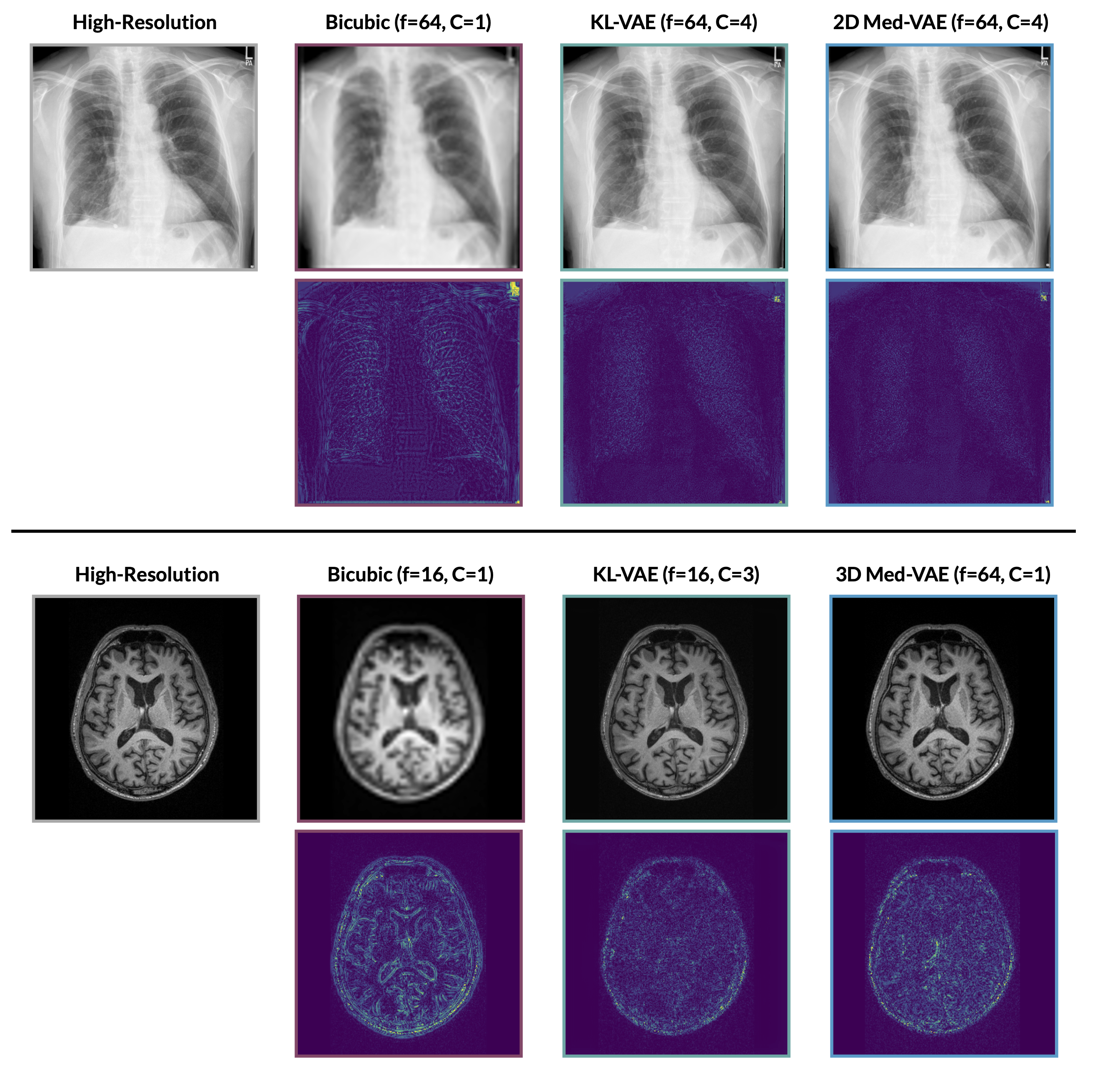}
\caption{\textbf{Qualitative examples of reconstructed medical images.} The top section provides qualitative examples of a reconstructed chest X-ray. The bottom section provides qualitative examples of a reconstructed brain MRI slice. Residual figures show pixel-level differences between reconstructed images and original, high-resolution images; brighter colors represent larger differences.}
\label{fig:qualitative}
\end{figure}

For manual evaluations of reconstructed image quality, readers rated image fidelity for 2D MedVAE to be 2.8 points higher than bicubic interpolation averaged across the two downsizing factors. 2D MedVAE also better preserved clinically-relevant features (2.8 points). Artifacts (e.g. blurring, hallucinations) were more frequent in interpolated images (2.6 points), which severely suffered from blurring artifacts with increasing downsizing factors. In summary, our results suggest that 2D MedVAE better preserves diagnostic features than interpolation. In Figure \ref{fig:qualitative}, we provide qualitative examples of a reconstructed chest X-ray and a reconstructed T1-weighted brain MRI slice.

\subsection{Ablations}

We analyze the effects of each stage of training on latent representation quality in Appendix Table~\ref{table:ablations2d} and Appendix Table~\ref{table:ablations3d}. 

We further ablate the inclusion of the embedding consistency loss term in the Stage 1 training procedure. We find that the embedding consistency loss term helps improve reconstructed image quality, particularly at lower compression factors. For instance, at a compression factor of $f=16$, Stage 1 training without the embedding consistency loss term achieves a PSNR of $37.27_{\pm 0.08}$ and an MS-SSIM of $0.992_{\pm 0.0}$ on mammograms. In comparison, Stage 1 training with the embedding consistency loss term achieves a PSNR of $37.57_{\pm 0.08}$ and an MS-SSIM of $0.993_{\pm 0.0}$, as shown in Table \ref{table:perceptualid}.
\clearpage
\section{Discussion}

High-resolution medical images can result in large data storage costs and increased or intractable computational complexity for trained models. As the volume of data stored by hospitals continues to increase and large-scale foundation models become more commonplace, methods for inexpensively storing and efficiently processing high-resolution medical images become a critical necessity. In this work, we aim to address this need by introducing MedVAE, a family of 6 large-scale autoencoders for medical images developed using a novel two-stage training procedure. MedVAE encodes high-resolution medical images as downsized latent representations. We demonstrate with extensive evaluations that (1) downsized latent representations can effectively replace high-resolution images in CAD pipelines while maintaining or exceeding performance, (2) downsized latent representations reduce storage requirements (up to 512x) and improve downstream efficiency (up to 70x in model throughput) when compared to high-resolution input images, and (3) reconstructed images effectively preserve relevant features necessary for clinical interpretation by radiologists.

Several prior works have introduced powerful autoencoders capable of generating downsized latents for images. In particular, recent work on latent diffusion models has involved the development of several large-scale autoencoders, such as VQ-GANs and VAEs, trained on eight million natural images~\cite{rombach2022high,kingma2013vae,esser2021taming,openimages}; downsized latents generated by these models were shown to capture relevant spatial structure as well as improve efficiency of downstream diffusion model training~\cite{rombach2022high}. However, recent works have demonstrated that models trained on natural images often generalize poorly to medical images due to significant distribution shift~\cite{guan2022,van2023exploring,chambon2022adapting}, suggesting that existing natural image autoencoders may not be well-suited for the complexity of the medical image domain. Our evaluations on both latent representations and reconstructed images support this point, demonstrating that existing large-scale natural image autoencoders consistently underperform our domain-specific medical image autoencoders. These findings demonstrate the need for domain-specific models capable of understanding complex and fine-grained patterns across diverse imaging modalities and anatomical regions.

Our work aims to reduce computational costs associated with automated medical image interpretation by proposing the use of training datasets comprised of downsized MedVAE latent representations rather than high-resolution medical images. For instance, given a chest X-ray training dataset with images of size $1024 \times 1024$ with 1 channel, our 2D MedVAE model with $f=64$ and $C=1$ can generate downsized latent representations of size $128 \times 128$ with 1 channel, contributing to substantial downstream efficiency and storage benefits. We demonstrate with eight CAD tasks that latent representations do not result in the loss of clinically-important information; at a 2D downsizing factor of $f=16$ and a 3D downsizing factor of $f=64$, we observe equivalent or better performance than high-resolution images with substantial improvements over multiple existing downsizing methods. MedVAE models can also generalize beyond the images included in the training set, as shown by performance on 2D musculoskeletal X-rays and 3D spine CTs. Importantly, the efficiency benefits of using latent representations are significant; in particular, using latent representations can contribute to large increases in batch sizes, which can be particularly useful in the modern era of self-supervised foundation models that rely heavily on the use of large batch sizes during training. 

The MedVAE autoencoder family includes two 3D autoencoders that are explicitly designed to downsize 3D medical imaging modalities (e.g. CT, MRI), a previously underresearched setting. Our results demonstrate that at a 3D downsizing factor of $f=64$, the volumetric latent representations generated by 3D MedVAE are substantially higher quality than those generated by stitching together 2D slices downsized using 2D baselines. This suggests that 3D autoencoders can better capture clinically-important volumetric patterns, such as fractures that span multiple slices. Efficiency benefits in the 3D setting are also notable, particularly since training downstream CAD models on high-resolution 3D volumes is often computationally expensive or intractable. At significantly higher downsizing factors ($f=512$), we observe the benefits of 3D autoencoder training to be less pronounced, suggesting that users will need to carefully consider the tradeoffs between latent representation quality and desired downstream efficiency when selecting a MedVAE model.

In addition to generating high-quality latent representations, MedVAE models also include a trained decoder, which can reconstruct the original high-resolution image from the downsized latent. This is a particularly useful capability in the medical imaging domain, since high-resolution images are necessary for effective clinical interpretation by radiologists. We demonstrate with a reader study consisting of three radiologists that reconstructed images can effectively preserve clinically-relevant signal needed for diagnoses; in this setting, fine-grained fractures in chest X-rays were preserved through the encoding and decoding process.

Our study presents several opportunities for future work. First, additional research into model architectures, data augmentation approaches, and training strategies would be useful for building effective downstream CAD models that can learn from latent representations. In addition, the batch size and efficiency benefits afforded by latent representations raise the possibility of training large-scale foundation models using downsized latent representations. Whereas foundation models traditionally require significant computational resources and training time, utilizing downsized latent representations that preserve diagnostic features can greatly accelerate model training, particularly in resource-constrained settings. Future work can explore foundation model performance and scaling laws in this context. Finally, future work can explore additional autoencoder training strategies to better preserve clinically-relevant features at high downsizing factors. 

Overall, our work demonstrates the potential that large-scale, generalizable autoencoders hold in addressing critical storage and efficiency challenges in the medical domain. 

\clearpage
\section{Methods}
\subsection{Background}
In this section, we provide background information on autoencoders. 

2D autoencoding methods can be formulated as follows. We begin with a training dataset $\mathcal{D} = \{x_i\}_{i=1}^N$ consisting of $N$ high-resolution input images $x_i \in \mathcal{X}$. Each high-resolution image $x_i$ has dimensions $H \times W$ with $B$ channels, which can be expressed as $x_i \in \mathbb{R}^{H \times W \times B}$. An autoencoding method learns an encoding function $g: \mathcal{X} \rightarrow \mathcal{Z}$, where $\mathcal{Z}$ represents a low-dimensional latent space and $z_i \in \mathcal{Z}$ represents the downsized latent representation corresponding to the input $x_i$. Let $f$ represent the downsizing factor applied to the 2D area of the image; then, the latent representation $z_i$ can be expressed as $z_i \in \mathbb{R}^{(H/(\sqrt{f}) \times (W/\sqrt{f}) \times C}$, where $C$ is a pre-specified number of latent channels. Autoencoding methods also learn a decoding function $h: \mathcal{Z} \rightarrow \hat{\mathcal{X}}$, which reconstructs the image $\hat{x_i}$ from the latent representation $z_i$. The encoding and decoding functions $g$ and $h$ are optimized in an end-to-end manner with the goal of maximizing perceptual similarity between $x_i$ and $\hat{x_i}$.

3D autoencoding methods follow a similar formulation, where each image $x_i$ represents a 3D volume with dimensions $H \times W \times S$ with $B$ channels. Here, the downsizing factor $f$ is applied to the 3D volume of the image; as a result, the latent representation $z_i$ can be expressed as $z_i \in \mathbb{R}^{(H/(\sqrt[3]{f}) \times (W/\sqrt[3]{f}) \times (S/(\sqrt[3]{f}) \times C}$, where $C$ is a pre-specified number of latent channels. 

\subsection{Curating a large-scale training dataset}
We first collect a large-scale, open-source training dataset $\mathcal{D}$ for training medical image autoencoders. We incorporate diverse modalities and anatomical features in order to ensure that trained autoencoders gain proficiency in processing the wide variety of diagnostic features that occur in medical images. Our dataset consists of 1,021,356 2D images and 31,374 3D images obtained from 19 multi-institutional, open-source datasets.

2D images include chest X-rays and FFDMs, selected because (a) chest X-rays are well-studied with large amounts of publicly-available data and (b) FFDMs are a challenging class of images due to large dimensions and the presence of fine-grained features critical for diagnoses (e.g. microcalcifications). We collect images from two chest X-ray datasets and six FFDM datasets~\cite{johnson2019mimic,feng2021candid,jeong2022emory,sorkhei2021csaw,rsnamammo,nguyen2022vindrmammo,moreira2012inbreast,cai2023online}. 

3D images include head MRIs, knee MRIs, and high-resolution whole-body (head, neck, abdomen, chest, lower limb) CTs. We selected these datasets since (a) head MRIs/CTs are a commonly obtained examination, and (b) high-resolution CTs tend to contain subtle features and consume large amounts of storage. These images were curated from four T1- and T2-weighted head MRI datasets (14,296), one knee MRI dataset (3,564), two head/neck CT datasets (10,156), two whole-body CT datasets (1,434), and two chest CT datasets (1,924)~\cite{jack2008alzheimer,dagley2017harvard,insel2020a4,lamontagne2019oasis,bien2018deep,hooper2021impact,chilamkurthy2018development,wasserthal2023totalsegmentator,ji2022amos,armato2011lung,stanfordaimi_coca_2024}.

\subsection{Training autoencoders for medical images}
In this section, we discuss our two-stage approach for training generalizable autoencoders for medical images. Motivated by prior work on natural images~\cite{rombach2022high}, we elect to use variational autoencoders (VAEs) as our backbone. In the first stage of training, we optimize for reconstruction quality by maximizing perceptual similarity between the input image $x$ and the reconstructed image $\hat{x}$. Whereas existing works train autoencoders solely using this approach, the medical image domain introduces the added complexity of subtle, fine-grained features required for clinical interpretation of images; thus, we introduce a second stage of training, where the latent representation space $\mathcal{Z}$ is refined with continued fine-tuning. Our approach is intended to explicitly preserve diverse clinically-relevant features in both latent representations and reconstructed images. In total, the MedVAE family includes four 2D VAEs and two 3D VAEs trained with various downsizing factors.

\textbf{Stage 1: Training Base Autoencoders} (Fig.~\ref{fig:method}a). 
We begin by performing base training of the autoencoders using the collected 2D images in order to optimize the quality of reconstructions $\hat{x}$. In line with prior work~\cite{rombach2022high}, each MedVAE autoencoder learns an encoder and decoder (corresponding to functions $g$ and $h$) end-to-end using a fully convolutional VAE. Each MedVAE autoencoder accepts single-channel, high-resolution medical images $x_i$ as input, applies function $g$ to transform the input to a downsized latent representation $z_i$, and then applies function $h$ to reconstruct the original image $\hat{x_i}$. MedVAE models are characterized by two hyperparameters: $f$, which represents the downsizing factor applied to the 2D area of the input image, and $C$, which describes the number of channels included in the latent representation. For instance, given an input image $x_i$ of size $H \times W \times 1$, a MedVAE model with $f = 16$ and $C = 3$ would generate a latent representation $z_i$ of size $(H/4) \times (W/4) \times 3$, downsizing the image area by 16x and adding two additional channels. The reconstructed image $\hat{x_i}$ would be of size $H \times W \times 1$.

In order to learn functions $g$ and $h$, the VAE is trained to maximize the similarity between $x_i$ and $\hat{x_i}$ using a perceptual loss term~\cite{lpips} and a patch-based adversarial objective~\cite{isola2018patchgan}. Additionally, in order to ensure preservation of clinically-relevant features within the reconstructed image, we introduce a domain-specific embedding consistency loss based on BiomedCLIP, a pretrained vision-language foundation model trained on a large corpus of paired medical image-text data~\cite{zhang2023biomedclip}. During training, we apply an $L_2$ penalty between BiomedCLIP embeddings corresponding to the input image $x_i$ and the reconstructed image $\hat{x_i}$. This loss function is inspired by prior work on developing autoencoders for chest X-rays~\cite{lee2023llmcxr}. Finally, in addition to the loss functions listed above, a KL-divergence penalty is applied to the latent sample in order to pull latents towards a standard normal; the penalty is assigned a low weight of 1e-6. 

We use the above loss functions and the curated dataset of one million 2D images to train the following four base autoencoders, trained across various downsizing factors and latent channels. Implementation details for each base model is described below:
\begin{itemize}
\item \textbf{2D Base Autoencoder (Stage 1) with $f=16$ and $C=1$}: This autoencoder yields latent representations $z_i$ of size $(H/4) \times (W/4) \times 1$. Stage 1 training is performed from scratch. The VAE is trained solely with the perceptual loss, the KL-divergence penalty, and the BiomedCLIP embedding consistency loss for the first 3125 steps; then, the patch-based adversarial objective is applied. We train for 100K steps using 8 NVIDIA A100 GPUs and a batch size of 32. 
\item \textbf{2D Base Autoencoder (Stage 1)  with $f=16$ and $C=3$}: This autoencoder yields latent representations $z_i$ of size $(H/4) \times (W/4) \times 3$. We first initialize the VAE with weights from a previously-developed natural image autoencoder (KL-VAE)~\cite{rombach2022high}. Then, we perform Stage 1 training using LoRA~\cite{hu2021lora} with rank=4 applied to all 2D convolutional layers. We train with all four loss functions for 50k steps using 8 A100 GPUs and a batch size of 32. 
\item \textbf{2D Base Autoencoder (Stage 1)  with $f=64$ and $C=1$}: This autoencoder yields latent representations $z_i$ of size $(H/8) \times (W/8) \times 1$. Stage 1 training is performed from scratch. The VAE is trained solely with the perceptual loss, the KL-divergence penalty, and the BiomedCLIP embedding consistency loss for the first 3125 steps; then, the patch-based adversarial objective is applied. We train for 100K steps using 8 NVIDIA A100 GPUs and a batch size of 32. 
\item \textbf{2D Base Autoencoder (Stage 1)  with $f=64$ and $C=4$}: This autoencoder yields latent representations $z_i$ of size $(H/8) \times (W/8) \times 4$. We first initialize the VAE with weights from a previously-developed natural image autoencoder (KL-VAE)~\cite{rombach2022high}. Then, we perform Stage 1 training using LoRA~\cite{hu2021lora} with rank=4 applied to all 2D convolutional layers. We train with all four loss functions for 50k steps using 8 A100 GPUs and a batch size of 32. 
\end{itemize}

\textbf{Stage 2: Preserving Clinically-Relevant Features Across Modalities} (Fig.~\ref{fig:method}b). After performing base training of the autoencoders using the collected 2D images, we introduce a second stage of training intended to further refine the latent space such that clinically-relevant features are preserved across various modalities. 

In the context of 2D imaging modalities, the second training stage takes the form of a lightweight fine-tuning procedure designed to maximize consistency in clinically-relevant features between the input image and the latent representation. Our key insight here is that image embeddings generated by BiomedCLIP~\cite{zhang2023biomedclip} can effectively capture clinically-relevant features in 2D medical images, suggesting utility as a guidance mechanism during training\footnote{We use the \texttt{BiomedCLIP-PubMedBERT\_256-vit\_base\_patch16\_224} model available on HuggingFace at \href{https://huggingface.co/microsoft/BiomedCLIP-PubMedBERT_256-vit_base_patch16_224}{https://huggingface.co/microsoft/BiomedCLIP-PubMedBERT\_256-vit\_base\_patch16\_224}.}. We freeze all parameters in the encoder and decoder of the VAE. During training, the input image $x_i$ is passed through the frozen VAE encoder to generate the latent representation $z_i$; then, $z_i$ is passed through a series of lightweight, trainable projection layers, which yield an output representation $\Bar{z_i}$ with the same size as $z_i$. Let the function $b(\cdot)$ represent the BiomedCLIP embedding function. We optimize the projection layer weights using a domain-specific embedding consistency loss, which takes the form of an $L_2$ loss between $b(x_i)$ and $b(\Bar{z_i})$. All downstream evaluations of latent representation quality are performed with the projected latent $\Bar{z_i}$. We perform Stage 2 training using the curated 2D training dataset with one million images. Our procedure yields four 2D MedVAE autoencoders with various downsizing factors and number of latent channels:
\begin{itemize}
\item \textbf{2D MedVAE with $f=16$ and $C=1$}: The projection layers generate $\Bar{z_i}$ of size $(H/4) \times (W/4) \times 1$. Stage 2 training is performed for 50K steps using 8 NVIDIA A100 GPUs and a batch size of 32. 
\item \textbf{2D MedVAE with $f=16$ and $C=3$}: The projection layers generate $\Bar{z_i}$ of size $(H/4) \times (W/4) \times 3$. Stage 2 training is performed for 50K steps using 8 NVIDIA A100 GPUs and a batch size of 32. 
\item \textbf{2D MedVAE with $f=64$ and $C=1$}: The projection layers generate $\Bar{z_i}$ of size $(H/8) \times (W/8) \times 1$. Stage 2 training is performed for 60K steps using 8 NVIDIA A100 GPUs and a batch size of 32. 
\item \textbf{2D MedVAE with $f=64$ and $C=4$}: The projection layers generate $\Bar{z_i}$ of size $(H/8) \times (W/8) \times 4$. Stage 2 training is performed for 50K steps using 8 NVIDIA A100 GPUs and a batch size of 32. 
\end{itemize}

In the context of 3D imaging modalities (e.g. CT scans, MRIs), the second training stage involves lifting the 2D VAE architecture to 3D using a kernel centering inflation strategy~\cite{zhang2022adapting}; we then continue training with 3D images. We note here that using external 2D medical foundation models like BiomedCLIP to enforce feature consistency is inadequate for 3D settings. As a result, we instead implement a training procedure focused on maximizing perceptual similarity, analogous to 2D stage 1 training. We train the 3D autoencoders using random cubic patches of size $64 \times 64 \times 64$. The perceptual loss and the patch-based adversarial objective are calculated per-slice, with the final loss term computed as the mean across all slices in the volume. Following such a training strategy, a 3D MedVAE model with $f = 64$, $C = 1$, and input image $x_i$ of size $H \times W \times S \times 1$ would generate a latent representation $z_i$ of size $(H/4) \times (W/4) \times (S/4) \times 1$, downsizing the volume by 64x. We perform Stage 2 training using the curated dataset of 31,374 3D images. Our procedure yields two 3D MedVAE autoencoders across various downsizing factors: 
\begin{itemize}
\item \textbf{3D MedVAE with $f=64$ and $C=1$}: The latent representations $z_i$ are of size $(H/4) \times (W/4) \times (S/4) \times 1$. We initialize the VAE with weights from 2D Base Autoencoder (Stage 1) with $f=16$ and $C=1$. We then train the VAE for 35K steps using 4 NVIDIA A6000 GPUs and a batch size of 32.
\item \textbf{3D MedVAE with $f=512$ and $C=1$}: The latent representations $z_i$ are of size $(H/8) \times (W/8) \times (S/8) \times 1$. We initialize the VAE with weights from 2D Base Autoencoder (Stage 1) with $f=64$ and $C=1$. We then train the VAE for 140K steps using 1 NVIDIA A6000 GPU and a batch size of 8. Both 3D MedVAEs are trained for the same number of steps when accounting for batch size. 
\end{itemize}

We analyze the effects of each stage of training on latent representation quality in Appendix Table~\ref{table:ablations2d} and Appendix Table~\ref{table:ablations3d}.

\subsection{Evaluating latent representations}
We evaluate the quality of latent representations $z$ with a set of eight clinically-relevant CAD tasks, which directly evaluate the preservation of clinically-relevant features in 2D and 3D images (Appendix Table~\ref{table:clssummary}). For each CAD task, we measure the difference in classification performance between models trained using latent representations and those trained using original, high-resolution images; this serves as an indicator of latent quality by directly measuring the retention of important diagnostic features. These evaluations also provide insights into potential performance gains afforded by training downstream models directly on MedVAE latent representations rather than high-resolution images.

Below, we provide implementation details for each 2D CAD task.
\begin{enumerate}
    \item  \textbf{Malignancy Detection:} We evaluate the quality of FFDM latent representations on a binary malignancy detection task, which involves predicting the presence or absence of a malignancy. We use images from the Chinese Mammography Dataset (CMMD), which includes a total of 5202 deidentified FFDMs from 1775 patients~\cite{cai2023online, cmmddata}. CMMD includes labels indicating the presence of masses and calcifications as well as biopsy-confirmed labels indicating benign and malignant findings. We assigned 80\% of patients to the training set (1420 patients with 2982 images) and the remaining 20\% to the test set (355 patients with 762 images). The average size of an FFDM after preprocessing was $1999.2 \times 793.9 \times 1$. In order to maintain consistent sizing, we downsized each FFDM to $1024 \times 512 \times 1$ using bicubic interpolation. 
    \item \textbf{Calcification Detection:} We evaluate the quality of FFDM latent representations on a binary calcification detection task, which involves identifying the presence or absence of breast calcifications. We use the CMMD dataset, described in detail above~\cite{cmmddata, cai2023online}. We preprocessed the CMMD dataset by assigning 80\% of patients to the training set (1420 patients with 4156 images) and 20\% of patients to the test set (355 patients with 1046 images). 
    \item \textbf{BI-RADS Classification:} We evaluate the quality of FFDM latent representations on Breast Imaging Reporting and Data System (BI-RADS) classification. We use images from the VinDR-Mammo dataset, which includes a total of 20,000 deidentified FFDMs from 5000 studies collected from Hanoi Medical University Hospital and Hospital 108 in Vietnam~\cite{nguyen2022vindrmammo}. BI-RADS scores evaluate the likelihood of cancer on an integer scale from 0 to 6~\cite{nguyen2022vindrmammo}. We use the provided data splits for VinDR-Mammo, which assign 16,000 images to the training set and 4000 images to the test set. There are no images with BI-RADS scores of 0 or 6. The average size of an FFDM after preprocessing was $2607.3 \times 948.6 \times 1$. In order to maintain consistent sizing across the dataset, we downsized each X-ray to $1024 \times 512 \times 1$. 
    \item \textbf{Bone Age Prediction:} We evaluate the quality of musculoskeletal X-ray latent representations on a bone age prediction task. We use images from the RSNA Bone Age dataset, which includes 14,036 hand radiographs collected from Children’s Hospital Colorado and Lucile Packard Children’s Hospital at Stanford University~\cite{rsnaboneage}. We use the provided data splits for the RSNA Bone Age dataset, which assign 12,611 images to the training set and 1425 images to the test set.  The average size of a musculoskeletal X-ray after preprocessing was $1665.4 \times 1319.8 \times 1$. In order to maintain consistent sizing across the dataset, we downsized each X-ray to $1024 \times 1024 \times 1$.
    \item \textbf{Pediatric Wrist Fracture Detection:} We evaluate the quality of musculoskeletal X-ray latent representations on a binary wrist fracture detection task. We use images from the GRAZPEDWRI-DX dataset, which includes a total of 20,327 deidentified images from 6,091 patients collected at University Hospital Graz in Austria~\cite{Nagy2022wristfrac}. We preprocessed the GRAZPEDWRI-DX dataset by first using provided labels to remove all samples with metal hardware and casts, which may exhibit spurious correlations with the target labels. We then assigned 75\% of patients to the training set (4281 patients with 10,511 images) and the remaining 25\% to the test set (1428 patients with 3602 images). The average size of a musculoskeletal X-ray after preprocessing was $987.8 \times 537.7 \times 1$. In order to maintain consistent sizing across the dataset, we resized each X-ray to $1024 \times 512 \times 1$.
\end{enumerate}

We perform each 2D CAD task listed above using a pretrained HRNet\_w64 neural network implemented in the \texttt{timm} Python package~\cite{wang2020hrnet,timm}. HRNets are a type of convolutional neural network adapted for classification of high-resolution images. We preprocess latent representations by applying the mean operation across the channel dimension if more than one channel is present. We train the HRNet on 2 A100 GPUs using supervised linear probing with one output class. We train for 100 epochs using a batch size of 256, an AdamW optimizer~\cite{adamw} with an initial learning rate of 1e-4, and cross-entropy loss. Classification performance is measured on the test set using the final model checkpoint. We report AUROC for binary classification tasks and Macro AUROC for multi-class classification tasks. 

Below, we provide implementation details for each 3D CAD task.
\begin{enumerate}
    \item \textbf{Spine Fracture Detection:} We evaluate the quality of Spine CT latent representations on a binary spine fracture detection task. We use images from the VerSe 2019 dataset~\cite{loffler2020vertebral}, which includes 160 high-resolution, 1-mm isotropic or in sagittal 2-mm to 3-mm series of 1-mm in-plane resolution, spine CT images. The training, validation, and testing split (50/25/25) was maintained from the original dataset. The final size of a volume after preprocessing was $224 \times 224 \times 160$. 
    \item \textbf{Head Fracture Detection:} We evaluate the quality of head CT latent representations on a binary head fracture detection task. We use images from the CQ500 dataset~\cite{chilamkurthy2018development}, which includes 378 head CT images. This dataset was curated by the Centre for Advanced Research in Imaging, Neurosciences, and Genomics (CARING) in New Delhi, India. Images were divided into training and testing sets following an 80/20 split. The final size of a volume after preprocessing was $224 \times 224 \times 44$. 
    \item \textbf{ACL and Meniscal Tear Detection:} We evaluate the quality of knee MRI latent representations on a binary ACL or meniscal tear detection task. We use images from the MRNet dataset~\cite{bien2018deep}, which includes 1250 sagittal knee MRI scans performed at Stanford University Medical Center between 2001-2012. A positive label in this context may indicate the presence of an ACL tear, a meniscal tear, or both simultaneously. The dataset was split into a training and test set (95/5). The final size of a volume after preprocessing was $56 \times 256 \times 256$.
\end{enumerate}

We perform each 3D CAD task listed above using the MONAI SEResNet-152~\cite{hu2018squeeze} architecture. We implemented a weighted sampling strategy for the head fracture detection and ACL and meniscal tear detection tasks due to class imbalance. We trained the SEResNet-152 on an A6000 GPU using supervised linear probing with 1 output class. We trained for 100 epochs with a batch size of 20 for latents, a batch size of 10 for the original images, an AdamW optimizer~\cite{adamw} with an initial learning rate of 1e-4, and binary cross-entropy loss. Classification performance (AUROC) is measured on the test set using the final model checkpoint.

\subsection{Evaluating reconstructed images}
We evaluate the quality of reconstructions $\hat{x}$ using both automated and manual perceptual quality evaluations. Perceptual quality assessments measure information loss resulting from the autoencoding process by comparing the original image to the reconstructed (decoded) image. These evaluations quantify the extent to which the encoding and subsequent decoding process retains relevant features.

For 2D images, we evaluate full-image perceptual quality on chest X-rays, FFDMs, and musculoskeletal X-rays; we also evaluate fine-grained perceptual quality on musculoskeletal X-rays. Chest X-rays are obtained from CANDID-PTX~\cite{feng2021candid} and MIMIC-CXR~\cite{johnson2019mimic}; FFDMs are obtained from RSNA Mammography~\cite{rsnamammo}, VinDR-Mammo~\cite{nguyen2022vindrmammo}, CSAW-CC~\cite{sorkhei2021csaw}, EMBED~\cite{jeong2022emory}, CMMD~\cite{cai2023online}, and INBreast~\cite{moreira2012inbreast}; musculoskeletal X-rays are obtained from GRAZPEDWRI-DX~\cite{Nagy2022wristfrac}. We compute two standard perceptual quality metrics: PSNR and MS-SSIM. For 2D fine-grained perceptual quality evaluations, we extract 7677 images containing fractures from GRAZPEDWRI-DX, and we use bounding boxes provided by the authors to isolate the region of the fracture~\cite{Nagy2022wristfrac}. We then compute PSNR scores on these regions.

For 3D full-volume perceptual quality evaluations, we evaluate full-image perceptual quality on head MRIs, head CTs, abdomen CTs, whole-body CTs, lung CTs, and knee MRIs. Head MRIs are obtained from Alzheimer's Disease Neuroimaging Initiative (ADNI)~\cite{jack2008alzheimer}, Harvard Aging Brain Study (HABS)~\cite{dagley2017harvard}, A4 dataset~\cite{insel2020a4}, and Open Access Series of Imaging Studies (OASIS) brain dataset~\cite{lamontagne2019oasis}; head CTs are obtained from CQ500~\cite{chilamkurthy2018development}; whole-body CTs are obtained from TotalSegmentator dataset~\cite{wasserthal2023totalsegmentator}; abdomen CTs are obtained from the Abdominal Multi-Organ Segmentation (AMOS) dataset~\cite{ji2022amos}; lung CTs are obtained from LIDC-IDRI~\cite{armato2011lung}; and knee MRIs are obtained from MRNet~\cite{bien2018deep}. For each volume, a center crop of volume dimensions $160 \times 160 \times 160$ was extracted. For the AMOS and CQ500 datasets, the crop region was expanded to dimensions $320 \times 320 \times 160$ to include both soft-tissue and bony features. We compute two standard perceptual quality metrics: PSNR and MS-SSIM. 

For manual evaluations of reconstructed image quality, we perform a reader study with 3 radiologists. Each expert reader is presented with a pair of chest X-rays, consisting of an original high-resolution image $x$ on the left and a reconstructed image $\hat{x}$ on the right (Appendix Fig.~\ref{fig:readerstudyui}). A total of 50 unique chest X-rays with fractures, randomly sampled from CANDID-PTX, are selected and presented in a randomized order~\cite{feng2021candid}. The reader study poses three distinct questions on image fidelity, preservation of clinically-relevant features, and the presence of artifacts. Each question is scored based on a 5-point Likert scale ranging between -2 and 2. Below, we provide additional details on each of these questions: 
\begin{enumerate}
    \item \textbf{Image Fidelity:} This question aims to assess how closely the reconstructed CXR image resembles the original image in terms of image fidelity considering the overall similarity, level of detail preservation, and visual quality. A higher rating indicates a closer resemblance to the original image, while a lower rating implies a greater deviation or degradation.
    \item \textbf{Preservation of clinically-relevant features:} This question evaluates the extent to which the reconstructed chest X-rays image preserves the diagnostic information present in the original image given the clarity and visibility of anatomical structures, abnormalities, and other important diagnostic features. A higher rating indicates a greater preservation of diagnostic information, while a lower rating suggests a significant loss that may affect the accuracy of diagnosis.
    \item \textbf{Presence of Artifacts:} This question focuses on the presence and impact of artifacts in the reconstructed chest X-ray. Artifacts can include image distortions, noise, blurring, or other visual anomalies (ie. hallucinations) that are not present in the original image. A higher rating suggests less or no interference from artifacts, while a lower rating suggests a greater occurrence of artifacts.
\end{enumerate}

\subsection{Statistical analysis}
For latent representation evaluations, we report classification performance using AUROC, calculated using the \texttt{torchmetrics} library. We report mean and standard deviations across three runs with different random seeds. For automated perceptual quality evaluations on 2D images, we calculate PSNR and MS-SSIM on a random sample of 1000 images for each image type; we report mean and standard deviations across four runs with different random seeds. For automated perceptual quality evaluations on 3D images, we calculate PSNR and MS-SSIM on a single random sample of 100 images for each image type. For manual perceptual quality evaluations with expert readers, we report mean scores and 95\% confidence intervals across three readers.

\clearpage

\subsection*{Acknowledgments}
MV is supported by graduate fellowship awards from the Department of Defense (NDSEG), the Knight-Hennessy Scholars program at Stanford University, and the Quad program. AK is supported by graduate fellowships from the Knight-Hennessy Scholars program at Stanford University and the Tau Beta Pi society. RS was supported by the Rubicon fellowship of the Dutch National Research Council (NWO). AC is supported by NIH grants R01 HL167974, R01HL169345, R01 AR077604, R01 EB002524, R01 AR079431, P41 EB027060, AY2 AX000045, and 1AYS AX0000024-01; ARPA-H grants AY2AX000045 and 1AYSAX0000024-01; and NIH contracts 75N92020C00008 and 75N92020C00021. A.C. has provided consulting services to Patient Square Capital, Chondrometrics GmbH, and Elucid Bioimaging; is co-founder of Cognita; has equity interest in Cognita, Subtle Medical, LVIS Corp, Brain Key. CL is supported by NIH grants R01 HL155410, R01 HL157235, by AHRQ grant R18HS026886, and by the Gordon and Betty Moore Foundation. CL is also supported by the Medical Imaging and Data Resource Center (MIDRC), which is funded by the National Institute of Biomedical Imaging and Bioengineering (NIBIB) under contract 75N92020C00021 and through the Advanced Research Projects Agency for Health (ARPA-H). 

This research was funded, in part, by the Advanced Research Projects Agency for Health (ARPA-H). The views and conclusions contained in this document are those of the authors and should not be interpreted as representing the oﬃcial policies, either expressed or implied, of the U.S. Government.

\subsection*{Author contributions}
M.V., A.K., and R.S. designed the study, constructed models, and performed technical evaluations. R.S., C.B., and J.P. carried out the reader study. M.V., A.K., R.S., S.O., L.B., and P.C. collected data and analyzed model performance. All authors contributed to technical discussions and the drafting and revision of the manuscript. C.L. and A.C. supervised, funded, and guided the research.

\clearpage
\nolinenumbers
\bibliographystyle{plain} 
\bibliography{main}      
\clearpage
\section*{\hspace{-0.5em}Appendix}
\captionsetup[figure]{labelformat=default, labelsep=colon, name=Appendix Figure}
\captionsetup[table]{labelformat=default, labelsep=colon, name=Appendix Table}
\renewcommand{\thefigure}{\arabic{figure}}

\begin{figure*}[ht]
  \includegraphics[width=\columnwidth]{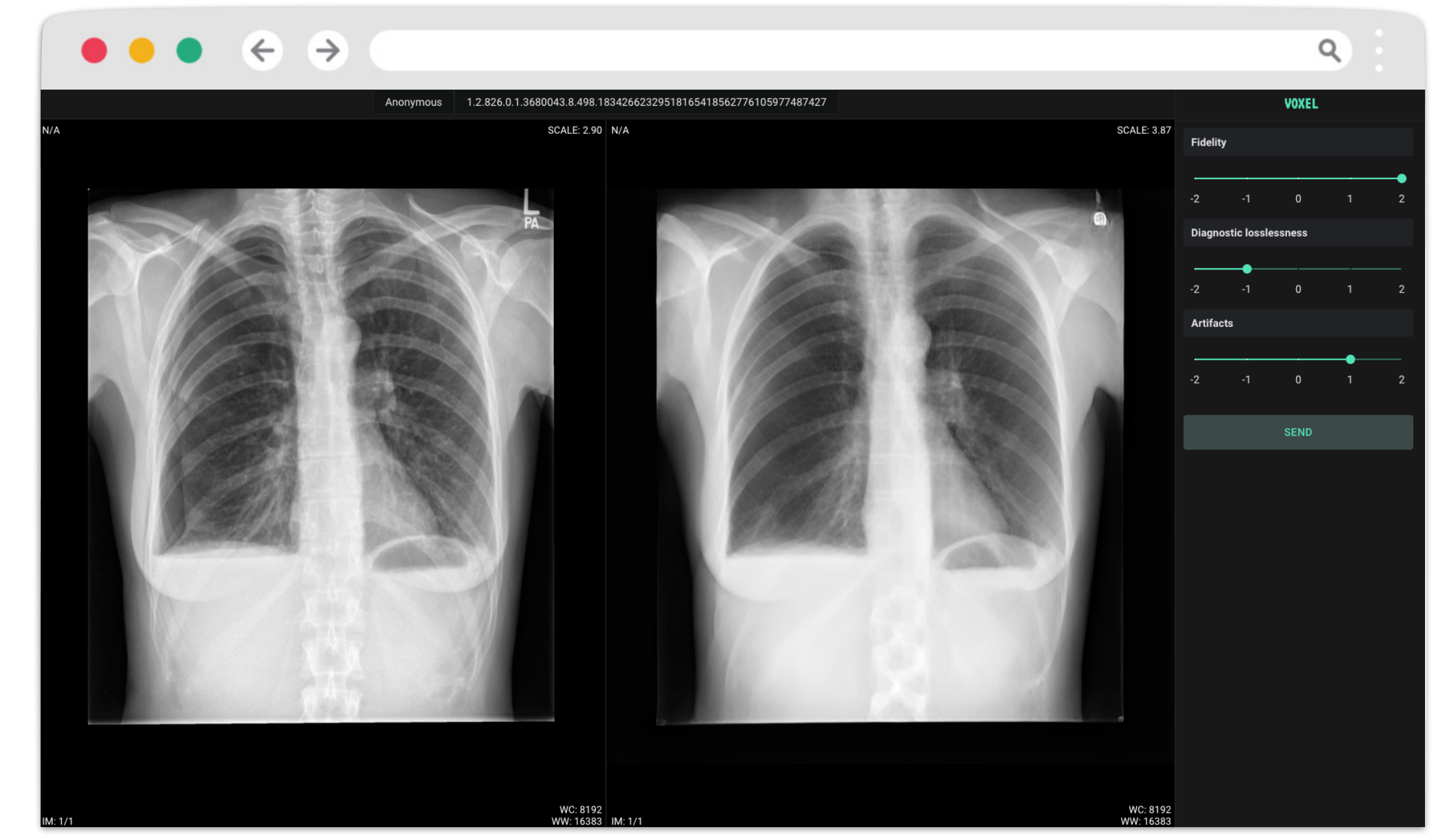}
  \caption{\textbf{Reader study user interface}. Expert readers score each reconstructed chest x-ray with respect to image fidelity, preservation of clinically-relevant features, and the presence of artifacts. Each expert reader is presented with a pair of chest X-rays, consisting of an original high-resolution image $x$ on the left and a reconstructed image $\hat{x}$ on the right. Readers are blinded to both the method and the downsizing factor used to generate the reconstructed image. }
  \label{fig:readerstudyui}
\end{figure*}

\begin{table*}[t]
\begin{center}
{\renewcommand{\arraystretch}{1.2}%
{
\resizebox{\linewidth}{!}{
\begin{tabular}{lccccccc}
\toprule
\textbf{Classification Task}
& \textbf{Image Dimensionality}
& \textbf{Num. Classes}
& \textbf{Dataset}
& \textbf{Modality}
& \textbf{Anatomy}
& \textbf{Num. Images}
\\ 
\midrule
Malignancy Detection & 2D & 2 & CMMD~\cite{cai2023online} & FFDM & Breast & 3744\\
Calcification Detection & 2D & 2 & CMMD~\cite{cai2023online} & FFDM & Breast & 5202\\
BI-RADS Classification & 2D & 5 & VinDR-Mammo~\cite{nguyen2022vindrmammo} & FFDM & Breast & 20,000 \\
Bone Age Prediction & 2D & 20 & RSNA Bone Age~\cite{rsnaboneage} & X-Ray & Hand & 14,036 \\
Wrist Fracture Detection & 2D & 2 &  GRAZPEDWRI-DX~\cite{Nagy2022wristfrac} & X-Ray & Wrist & 14,113 \\
Spine Fracture Detection & 3D & 2 &  VerSe~\cite{loffler2020vertebral} & CT & Spine & 160 \\
Head Fracture Detection & 3D & 2 &  CQ500~\cite{chilamkurthy2018development} & CT & Head & 378 \\
ACL \& Meniscal Tear Detection & 3D & 2 &  MRNet~\cite{bien2018deep} & MRI & Knee & 1250  \\
\bottomrule
\end{tabular}
}
}
}
\end{center}
\caption{\textbf{Summary of CAD tasks used for evaluating latent representation quality}. We report the task name, number of classes associated with the task, the dataset name, imaging modality, anatomical features, and the number of images after preprocessing.}
\label{table:clssummary}
\end{table*}

\begin{table*}[t]
\centering
\resizebox{\linewidth}{!}
{
\begin{tabular}{ lcccccccc }
\toprule
\textbf{}
& \multicolumn{2}{c}{\textbf{}}
& \multicolumn{5}{c}{\textbf{AUROC} $\uparrow$}
& \textbf{}
\\
\cmidrule(l{3pt}r{0pt}){4-8}
\cmidrule(l{3pt}r{0pt}){9-9}
\textbf{Method}
& \textbf{$f$}
& \textbf{$C$}
& \small Malignancy
& \small Calcification
& \small BI-RADS
& \small Bone Age
& \small Wrist Fracture
& Avg.
\\ 
\midrule
\small High-Resolution & 1 & 1 & \textbf{66.1} & \textbf{62.4} & \textbf{63.4} & \textbf{80.2}  & \textbf{73.7} & \textbf{69.2}\\

\midrule
\small 2D Base Autoencoder (Stage 1) \hspace{0.5mm} & 16 & 3 &  58.7 & 60.5 & 58.0 & 72.0 & 64.3 & 62.7 \\
\small 2D MedVAE (Stage 2) \hspace{0.5mm} & 16 & 3 &   \textbf{66.1} &   \textbf{61.7} & \textbf{62.3} & \textbf{82.1}  &   \textbf{70.6} & \textbf{68.6}\\
\midrule

\small 2D Base Autoencoder (Stage 1) \hspace{0.5mm} & 64 & 4 &   63.4 &  54.4 & 58.6 & 65.7 & 61.9 & 60.8 \\
\small 2D MedVAE (Stage 2) \hspace{0.5mm} & 64 & 4  &  \textbf{64.9} &  \textbf{58.5} & \textbf{60.6} &\textbf{ 73.0} & \textbf{66.7} & \textbf{64.7}\\
\bottomrule
\end{tabular}
}
\caption{\textbf{Effect of each autoencoder training stage on 2D MedVAE latent representation quality.} We evaluate the effects of each stage of 2D MedVAE training on latent representation quality using five 2D CAD tasks.}
\label{table:ablations2d}
\vspace{-1mm}
\end{table*}

\begin{table*}[t]
\centering
\resizebox{0.8\linewidth}{!}
{
\begin{tabular}{ lcccccc }
\toprule
\textbf{}
& \multicolumn{2}{c}{\textbf{}}
& \multicolumn{3}{c}{\textbf{AUROC} $\uparrow$}
& \textbf{}
\\
\cmidrule(l{3pt}r{0pt}){4-6}
\cmidrule(l{3pt}r{0pt}){7-7}
\textbf{Method}
& \textbf{$f$}
& \textbf{$C$}
& \small Spine Fractures
& \small Skull Fractures
& \small Knee Injury
& Avg.
\\ 
\midrule
\small High-Resolution & 1 & 1 & \textbf{82.9} & \textbf{63.9} & \textbf{69.9} & \textbf{72.2}\\

\midrule
\small 2D Base Autoencoder (Stage 1) \hspace{0.5mm} & 64 & 1 & 76.1 & 36.6 & 65.0 & 59.2 \\
\small 3D MedVAE (Stage 2) \hspace{0.5mm} & 64 & 1 & \textbf{83.7} & \textbf{87.0} & \textbf{68.4} & {\textbf{79.7}} \\
\midrule

\small 2D Base Autoencoder (Stage 1) \hspace{0.5mm} & 512 & 1 & \textbf{72.5 } & 45.4 & \textbf{68.8} & 62.2 \\
\small 3D MedVAE (Stage 2) \hspace{0.5mm} & 512 & 1  & 72.0 & \textbf{49.1} & 58.2 & 59.8\\
\bottomrule
\end{tabular}
}
\caption{\textbf{Effect of each autoencoder training stage on 3D MedVAE latent representation quality.} We evaluate the effects of each stage of 3D MedVAE training on latent representation quality using three 3D CAD tasks. Since Stage 1 training exclusively involves 2D images, we evaluate this model on 3D tasks by stitching 2D latent representations together across slices such that the size of the 2D latent representation matches those generated by 3D models.}
\vspace{-1mm}
\label{table:ablations3d}
\end{table*}

\begin{table*}[ht]
\centering
{%
\begin{tabular}{ lcccccc }
\toprule
\textbf{}
& \multicolumn{2}{c}{\textbf{}}
& \multicolumn{3}{c}{\textbf{AUROC} $\uparrow$}
& \textbf{}
\\
\cmidrule(l{3pt}r{0pt}){4-6}
\cmidrule(l{3pt}r{0pt}){7-7}
\textbf{Method}
& \textbf{$f$}
& \textbf{$C$}
& \small Spine Fractures
& \small Skull Fractures
& \small Knee Injury
& Average
\\ 
\midrule
\small High-Resolution & 1 & 1  & \textbf{82.9$_{\pm2.2}$} & \textbf{63.9$_{\pm6.3}$} & \textbf{69.9$_{\pm0.6}$} & \textbf{72.2}\\

\midrule
\small 2D MedVAE  & 64 & 1  & 80.5$_{\pm4.9}$ & 57.4$_{\pm4.0}$ & 67.3$_{\pm3.6}$ &  68.4\\
\small 2D MedVAE & 64 & 3  & 78.6$_{\pm0.8}$ & 50.9$_{\pm19.5}$ & 60.9$_{\pm4.2}$ &  63.5\\
\small 3D MedVAE & 64 & 1  & \textcolor{blue}{\textbf{83.7$_{\pm2.8}$}} & \textcolor{blue}{\textbf{87.0$_{\pm7.3}$}} & \textbf{68.4$_{\pm2.4}$} & \textbf{79.7}\\
\midrule
\small 2D MedVAE & 512 & 1  & 65.9$_{\pm8.7}$ & \textbf{63.0$_{\pm1.1}$} & 55.9$_{\pm8.3}$ & \textbf{61.6} \\
\small 2D MedVAE  & 512 & 4  & \textbf{81.9$_{\pm1.2}$} & 17.1$_{\pm8.6}$ & 52.6$_{\pm1.9}$ & 50.5 \\
\small 3D MedVAE & 512 & 1  & 72.0$_{\pm3.8}$ & 49.1$_{\pm19.8}$ & \textbf{58.2$_{\pm1.7}$} & 59.8 \\
\bottomrule
\end{tabular}
}
\caption{\textbf{Comparing 2D MedVAE and 3D MedVAE on 3D CAD tasks.} We compare 3D MedVAE with 2D MedVAE models. For 2D MedVAE, we stitch 2D latent representations together across slices such that the size of the 2D latent representation matches those generated by the 3D model. Here, $f$ represents the downsizing factor applied to the 3D volume of the input image and $C$ represents the number of latent channels. The best performing models on each task are bolded. We highlight methods that perfectly preserve clinically-relevant features in \textcolor{blue}{blue}.}
\label{table:2d3dcad}
\vspace{-1mm}
\end{table*}

\begin{table*}[t]
\centering
\resizebox{\linewidth}{!}{
\begin{tabular}{lcccccccccccccc}
\toprule
\textbf{Method}
& \textbf{$f$}
& \textbf{$C$}
& \multicolumn{2}{c}{\textbf{Brain MRIs}}
& \multicolumn{2}{c}{\textbf{Head CTs}}
& \multicolumn{2}{c}{\textbf{Abdomen CTs}}
& \multicolumn{2}{c}{\textbf{TS CTs}}
& \multicolumn{2}{c}{\textbf{Lung CTs}}
& \multicolumn{2}{c}{\textbf{Knee MRIs}}
\\
\cmidrule(l{3pt}r{0pt}){4-5}
\cmidrule(l{3pt}r{0pt}){6-7}
\cmidrule(l{3pt}r{0pt}){8-9}
\cmidrule(l{3pt}r{0pt}){10-11}
\cmidrule(l{3pt}r{0pt}){12-13}
\cmidrule(l{3pt}r{0pt}){14-15}
& 
&
& \small PSNR $\uparrow$
& \small MS-SSIM $\uparrow$
& \small PSNR $\uparrow$
& \small MS-SSIM $\uparrow$
& \small PSNR $\uparrow$
& \small MS-SSIM $\uparrow$
& \small PSNR $\uparrow$
& \small MS-SSIM $\uparrow$
& \small PSNR $\uparrow$
& \small MS-SSIM $\uparrow$
& \small PSNR $\uparrow$
& \small MS-SSIM $\uparrow$
\\ 
\midrule
\small 2D MedVAE-Decoder & 64 & 1  & 28.88 & 0.978 & 35.01 & 0.997 & 31.47 & 0.983 & 29.96 & 0.981 & 27.54 & 0.965 & 27.04 & 0.992 \\
\small 3D MedVAE  & 64 & 1 & \textbf{29.52} &\textbf{ 0.983} & \textbf{39.03} & \textbf{0.999} & \textbf{36.61} & \textbf{0.993} & \textbf{31.35} & \textbf{0.987} & \textbf{28.79} & \textbf{0.975} & \textbf{28.25} &\textbf{ 0.994} \\
\midrule
\small 2D MedVAE-Decoder & 512 & 1 & 25.85 & 0.927 & 18.65 & 0.824 & 20.47 & 0.699 & 25.26 & 0.929 & 23.33 & 0.909 & 23.92 & 0.969 \\
\small 3D MedVAE  & 512 & 1 & \textbf{26.23} &\textbf{ 0.937} & \textbf{30.85} & \textbf{0.991} &\textbf{ 29.47} &\textbf{ 0.960} & \textbf{26.34} & \textbf{0.949} & \textbf{24.76} & \textbf{0.934} & \textbf{24.36} & \textbf{0.977} \\

\bottomrule
\end{tabular}
}
\caption{\textbf{Comparisons of 3D MedVAE and 2D MedVAE Decoder.} The 2D MedVAE-Decoder model performs downsizing on individual 2D slices, which are then stitched and interpolated together to form a latent representation of equivalent size to the 3D MedVAE model; we then perform fine-tuning of the decoder using our curated dataset of 3D volumes. We compare perceptual quality of reconstructed volumes across six 3D image types. Here, $f$ represents the downsizing factor applied to the 3D volume of the input image and $C$ represents the number of latent channels. The best performing models on each task are bolded.}
\label{table:decoder}
\end{table*}

\end{document}